\documentclass[runningheads]{svmult}

\usepackage{makeidx}   % allows index generation
\usepackage{graphicx}  % standard LaTeX graphics tool
\usepackage{subeqnar}  % subnumbers individual equations
\usepackage{multicol}  % used for the two-column index
\usepackage{physprbb}  % modified textarea for proceedings,
\makeindex             % used for the subject index
\usepackage{euscript}

\newcommand{\greeksym}[1]{{\usefont{U}{psy}{m}{n}#1}}
\newcommand{\umu}{\mbox{\greeksym{m}}}

\newcommand{\be}{\begin{equation}}

\def\lsim{\mathrel{\rlap{\lower4pt\hbox{\hskip1pt$\sim$}}
    \raise1pt\hbox{$<$}}}
\def\gsim{\mathrel{\rlap{\lower4pt\hbox{\hskip1pt$\sim$}}
    \raise1pt\hbox{$>$}}}
\def\lb{\langle}
\def\rb{\rangle}

\def\bfJ{\bf J}
\def\bfA{\bf A}
\def\bfa{\bf a}
\def\bbE{\overline {\bf E}}

\def\bw{\overline {\omega}}
\def\bv{\overline V}
\def\bB{\overline B}
\def\ts{\times}
\def\lb{\langle}
\def\rb{\rangle}
\def\curl{\nabla \ts}

\def\bfv{{\bf v}}

\def\bfV{{\bf V}}
\def\bfj{{\bf j}}

\def\bfe{{\bf e}}
\def\bfE{{\bf E}}
\def\bfw{{\bomega}}

\def\bfb{{\bf b}}

\def\bfB{{\bf B}}

\def\bbB{\overline{\bf B}}

\def\bbJ{\overline{\bf J}} 
\def\bbA{\overline{\bf A}} 
\def\bbE{\overline{\bf E}} 
\def\nb{\nabla}
\def\curl{\nb\ts}
\def\div{\nb\cdot}
\def\b0{b^{(0)}}
\def\v0{v^{(0)}}
\def\w0{\omega^{(0)}}
\def\bb0{\bfb^{(0)}}
\def\bv0{\bfv^{(0)}}
\def\bw0{\bfw^{(0)}}
\def\bj0{\bfj^{(0)}}

\def\be{\begin{equation}}
\def\ee{\end{equation}}
\def\lab#1{\label{#1}}
\def\lrp#1{\left(#1\right)}
\def\BV{{\bf V}}

\def\OV{\overline{\bf V}}
\def\E{{\bf E}}

\def\pref#1{(\ref{#1})}
\def\beq{\begin{eqnarray}}
\def\eeq{\end{eqnarray}}
\def\nn{\nonumber}
\def\nt{\nabla\times}
\def\OE{\overline{\bf E}}

\def\lra#1{\left\langle #1\right\rangle}
\def\bv{\bf v}
\def\OB{\overline{\bf B}}
\def\OA{\overline{\bf A}}

\def\cnt{\cdot\nabla\times}
\def\b{{\bf b}}
\def\ob{\overline{B}}

\def\la{\lambda}

\begin{document}
\title*
{Recent Developments in Magnetic Dynamo Theory}
\toctitle
{Recent Developments 
in  Magnetic Dynamo Theory}

% allows explicit linebreak for the table of content
%
%
\titlerunning
{Recent Developments in Magnetic Dynamo Theory}

% allows abbreviation of title, if the full title is too long
% to fit in the running head
%
\author{Eric G. Blackman  }

%\and David Grove\inst{1}
%\and Craig Chambers\inst{2}
%\and Kim~B.~Bruce\inst{2}
%\and Elsa Bertino\inst{1}}
%
\authorrunning{Eric G. Blackman}
% if there are more than two authors,
% please abbreviate author list for running head
%
%

\institute{Department of Physics and Astronomy, and Laboratory
for Laser Energetics, University of Rochester, 
Rochester NY 14627, USA
%\and Universit\'{e} de Paris-Sud,
%     Laboratoire d'Analyse Num\'{e}rique,
%     B\^{a}timent 425,\\
%     F-91405 Orsay Cedex, France
}

\maketitle              % typesets the title of the contribution

\medskip

\noindent (to appear in 
``Turbulence and Magnetic Fields in                
Astrophysics,'' eds. E. Falgarone and T. Passot, Springer 
Lecture Notes in Physics)

\begin{abstract}
\medskip

Two spectral regimes 
of magnetic field amplification in magnetohydrodynamic (MHD)
flows can be distinguished by the scale on which fields are amplified relative
to the primary forcing scale of the turbulence.
For field amplification at or below the forcing scale, 
the amplification can be called a ``small scale dynamo.'' 
For amplification at and above the forcing
scale the process can be called a ``large scale dynamo.''
$Non-local$ (in wave number) effects play a key
role in both 
the growth of the small scale field in non-helical turbulence and the
growth of large and small scale fields in helical turbulence.
Mean field dynamo (MFD)  theory represents a simple semi-analytic
way to get a handle on large scale field amplification in 
MHD turbulence. Helicity has long been known to be important
for large scale, flux generating, externally forced MFDs.
The extent to which such MFDs operate ``slow'' or ``fast'' 
(dependent or independent on magnetic Reynolds number) has been
controversial, but there has been recent progress. 
Simulations of $\alpha^2$ dynamos in a periodic box dynamo and their quenching 
can now be largely understood within a simplified dynamical 
non-linear paradigm in which the MFD growth equation
is supplemented by the total magnetic helicity evolution equation.
For $\alpha^2$ dynamos, the large scale field
growth is directly related to the large scale magnetic helicity growth.
Magnetic helicity conservation 
then implies that growth of the large scale magnetic helicity 
induces growth of small scale magnetic (and current) helicity 
of the opposite sign, which eventually 
suppresses the $\alpha$ effect driving the MFD growth.
Although the $\alpha^2$ MFD then becomes slow in the long time limit,
substantial large scale field 
growth proceeds in a kinematic, ``fast''  phase before non-linear 
asymptotic quenching of the ``slow'' phase applies.  
Ultimately, the MFD emerges as a process that transfers 
magnetic helicity between small and large scales.
How these concepts apply to more general
dynamos with shear, and  open boundary dynamos is a topic of ongoing research.
Some unresolved issues are  identified.
%Moreover, to understand whether astrophysical mean field dynamos 
%proceed slow or fast in light of understanding  slow box dynamos, 
%vertical stratification, vertical variation of dynamo coefficients,
%and radial shear must eventually be  included.
%Realistic boundary physics for dynamos 
%may play in important role: the observed shedding of current (and magnetic) 
%helicity in the sun may be helpful
%in ensuring a fast dynamo for all time.  
Overall, the following summarizes the most recent progress in mean-field
dynamo theory:

{\it For a closed turbulent flow,

the non-linear mean field dynamo,
 
is first fast and kinematic, then slow and dynamic,

and magnetic helicity transfer makes it so.}

\bigskip

\end{abstract}

\section{Small Scale vs. Large Scale Field Amplification}

A dynamo is  a process which 
exponentially amplifies or sustains magnetic energy 
in the presence of finite dissipation. 
In this paper I will focus on magnetohydrodynamic (MHD) dynamos,
where the only flux dissipating term in Ohm's law for the total
magnetic field is the resistive term.  

The simple definition of a dynamo given above
does not distinguish the scale on which 
the magnetic energy is sustained against turbulent
forcing, whether a net flux is produced 
in the spatial region of interest, or the nature
of the forcing (e.g. shear driven or isotropically forced).
It is helpful to distinguish between  ``small scale dynamos''
which describe field generation at or below the turbulent
forcing scale, and ``large scale dynamos'' 
which describe field generation on scales larger
than the forcing scale.  Both present their own set of problems.

For the Galaxy, 
supernovae dominate the nearly isotropic turbulent forcing, 
and although there is a range of
forcing scales, typically the dominant scale is 
$\sim 50-100$pc \cite{ruz}.
A cascade leads to a nearly Kolmogorov turbulent
kinetic energy spectrum. Faraday rotation and synchrotron polarization 
observations reveal  the presence of a random component of the Galactic
field, also with a dominant scale of  $\sim 50-100$ parsecs, and an 
ordered toroidal field on the scale of $\gsim 1$kpc 
%\cite{beck,Zweibel}. 

An important point about large scale field growth
is that regardless of long standing debates about whether the large
scale fields of the Galaxy  is primordial or produced in situ,
\cite{kulsrudetal,zweibel,ccm}
and regardless of similar debates about the origin of large
scale, jet producing, poloidal fields in accretion disks (e.g. \cite{lpp}) 
one thing we do know
is that the Sun, at least $must$ have a large scale dynamo operating because
the mean flux reverses sign every 11 years. If the flux
were simply flux frozen into the sun's formation from the protostellar
gas, we would not expect such reversals.

For both small and large scale dynamos, the richness 
and the complication of dynamo theory is the non-linearity of the MHD
equations. To really understand the theory, we need to 
understand the backreaction of the growing magnetic
field on the turbulence,  the saturated spectra of the
magnetic field, and the spectral evolution time scales.
Do theoretical and numerical calculations 
make predictions which are consistent with what is observed 
in astrophysical systems or not?  What are the limitations
of these predictions?

\subsection{Small Scale Dynamo}

The ``small scale dynamo'' describes magnetic field 
amplification on and below the turbulent forcing scale
(e.g. \cite{kazanstev,parker,zeldovich}).
%(Parker 1979; Zeldovich et al. 1983).
Field energy first builds up to equipartition with the kinetic
energy on the smallest scales \cite{kulsrud} because 
the growth time is the turnover time, the turnover times
are shorter for smaller scales, and the equipartition level
is lower for the small scales. 
The fast growth, seen 
in non-linear simulations, can be predicted 
analytically \cite{kazanstev,parker,zeldovich,kulsrud}.  
The approach to near equipartition is not controversial but the 
shape of the saturated small scale spectrum needs more discussion, 
particularly for  the Galaxy.  

Recent simulations of forced non-helical 
turbulence for magnetic Prandtl number ($Pr \equiv \nu_v/\nu_M$,
where $\nu_v$ is the viscosity and $\nu_M$ is the magnetic diffusivity)
satisfying $Pr \ge 1$ in a periodic box have shown 
that the field does not build up to anywhere 
near equipartition on the input scale of the turbulence
\cite{kida91,maroncowley}.  Rather, the field
piles up on the smallest scales. The interpretation 
is that the forcing scale velocity directly 
shears the field into folds or filaments
with length of order the
forcing scale but with cross field scale of order the resistive scale,
accounting for the spectral power at small scales.
(see  \cite{scmm} for a careful semi-analytic study of the $Pr >> 1$ case). 
Since the small scale field is amplified by shear directly from 
the input scale,  one can argue that there is a $non-local$ direct cascade.

How does the observed small scale field of the Galaxy compare
with the above results?  Since  observations show near 
equipartition of the field energy with the kinetic energy at the forcing scale
\cite{beck}, there is  a discrepancy between the observations
and the numerical simulation results for non-helical turbulence.

A possible resolution might arise from the
idealized problem,  explored in \cite{maronblackman}.  
When the turbulence is forced with sufficient  kinetic 
helicity $ \lb \bfv \cdot \curl\bfv\rb$,
the spectrum changes in an important way.  Figures 1 and 2 below 
come from 3-D MHD simulations in which kinetic energy is forced in 
a box, and the helical fraction of the kinetic forcing is varied
($f_h=1$ corresponds to maximal helical forcing).  
The forcing wave number is $k/2\pi=4.5$. 
For sufficiently large $f_h$, (determined by that for which
the kinematic $\alpha^2$ dynamo can grow)
the magnetic spectrum grows two peaks, 
one at $k/2\pi=1$ and one at the forcing scale.  
The kinetic helicity thus influences 
both the small and large scale field growth.

Whether these results apply to the Galaxy or protogalaxy 
is unclear since those systems 
have shear (see also \cite{vishniac}), 
boundaries, and stratification unlike the simulations
of \cite{maronblackman}.
But  the principle that  the large  $and$ small 
scale fields are both influenced by helicity has been demonstrated.

The emergence of two peaks motivates a two-scale approach  which is a great
simplification that will be exploited later.
The shift of the small scale  peak to the forcing scale 
is less understood than the generation of the large scale field.
I now discuss the latter.

\begin{figure}[]
\begin{center}
\includegraphics[width=.6\textwidth]{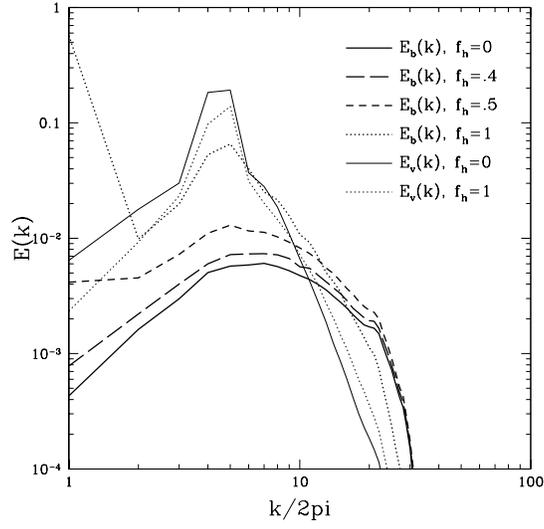}
\end{center}
\caption[]{
Saturated kinetic and magnetic energy spectra for successive values of
fractional helicity $f_h$. }
\label{helicity.spectra.ps}
\end{figure}

%------------------------------------------------------------------------------
%\vspace{-.1cm} \hbox to \hsize{ \hfill \epsfxsize8cm
%\epsffile{helicity.spectra.ps} \hfill } \noindent {\it Figure 1:
%Saturated kinetic and magnetic energy spectra for successive values of
%fractional helicity $f_h$. }

%-----------------------------------------------------------------------------

\begin{figure}[]
\begin{center}
\includegraphics[width=.6\textwidth]{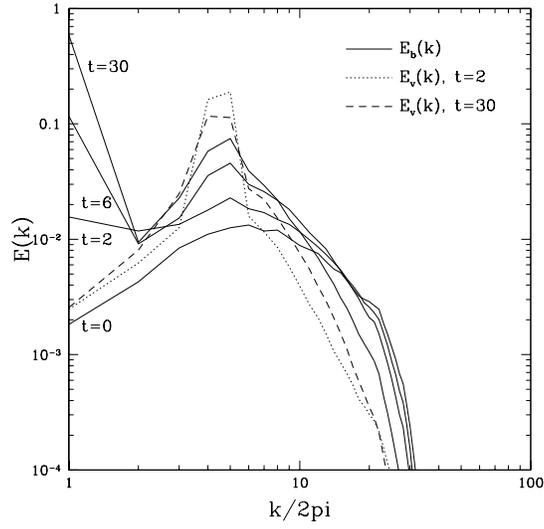}
\end{center}
\caption[]{Time sequence of kinetic and magnetic energy spectra for $f_h=1$}
\label{z338.ps}
\end{figure}

\subsection{Large Scale Dynamo}

The large scale field of the Galaxy \cite{beck,zweibel} 
seems to be of quadrapole mode and therefore the planar component of the
large scale field has the same sign across the mid-plane.  
The toroidal field  reverses on scales of a few kpc in radius 
In these  annuli, there appears to be a net toroidal flux
when integrated over the full height of the Galaxy.
If this inference continues to  survive future observations, 
the mechanism for field production must produce
a net toroidal flux in annuli that extend the full vertical disk
thickness, not just a net magnetic energy.  

A leading framework for understanding 
the in situ origin of large scale magnetic field energy and 
flux growth in galaxies and stars, 
and even for the peak in the large scale magnetic energy 
in the helically forced case of Figure 1  (see also \cite{b2001})
has been the mean field dynamo (MFD) theory \cite{ruz,parker,zeldovich,moffatt,krause}.  The theory appeals to some combination of helical
turbulence (leading to the $\alpha$ effect), differential rotation 
(the $\Omega$ effect),  
and turbulent diffusion (the $\beta$ effect) to exponentiate an initial seed mean magnetic field. 
Ref. \cite{steenbeck} developed a 
formalism for describing the concept \cite{parker} that helical turbulence  
can twist toroidal ($\phi$)  fields 
into the poloidal ($r,z$) direction, where they can be
acted upon by differential rotation to regenerate a powerful large scale
toroidal magnetic field.  The turbulent diffusion serves to
redistribute the flux so that inside the bounded volume of interest,
a net flux can grow. 

The formalism separates the total magnetic field into 
%$\B$ into 
a mean 
component $\OB$ and a fluctuating  component $\b$, and similarly for the
velocity field $\BV$.  The mean can be a spatial or ensemble average.  
The ensemble average is approximately equal to the spatial average  when there is a
scale separation between the mean and fluctuating scales.
%Otherwise, 
%the ensemble average need not correspond to an average that characterizes the
%specific realization for the object of interest. 
In reality, the scale separation is often weaker than the  
dynamo theorist desires, though a weak separation is also
helpful given the limited dynamic range of simulations.
I  proceed to consider spatial averages to simplify the discussion. 
(Please also note that non-helical large scale
field generation exists from the magnetorotational instability (MRI)
\cite{balbus} but I do not consider this here and focus
on externally forced systems.  Actually the
helical MFD dynamo may be operating even in systems with the MRI.
Also I do not consider the model of \cite{vishniac} here as that will
be covered elsewhere in this volume.)

The mean field $\OB$ satisfies the induction 
equation \cite{moffatt,steenbeck} 
\be
{\partial\OB\over \partial t} = -c\curl\bbE,
\lab{2.4a} 
\ee
where
\be
\bbE=-\lrp{\OV\times \OB}/c  - \lb\bfv\ts\bfb\rb/c+
\nu_M\nt\OB,
\lab{2.4aa} 
\ee
\be
\lb\bfv\ts\bfb\rb_i = \alpha_{ij}{\bB}_j-\beta_{ijk}\partial_j{\bB}_k
\lab{2.5a}
\ee
is the turbulent electromotive force (EMF), and $\nu_M=\eta c^2/4\pi$ is the magnetic
diffusivity defined with the resistivity $\eta$.
Here $\alpha_{ij}$ contains Parker's twisting (the $\alpha$ effect)
and $\beta_{ijk}$ contains the turbulent diffusivity. Ref. \cite{steenbeck}
calculated $\OE$ to first order in $\ob$ for isotropic
$\alpha_{ij}$ and $\beta_{ijk}$ and hence the pseudo-scalar 
and scalar dynamo coefficients $\alpha$ and 
$\beta$ respectively 
to zeroth order in $\ob$ from the statistics of the turbulence,
ignoring the Navier-Stokes equation.  
When the Navier-Stokes equation is not used, we speak of
the ``kinematic theory.'' 
Using the equation for the fluctuating field $\partial_t\bfb$,
 plugging it into $\lb\bfv\ts\bfb\rb$, the standard approach
gives a kinematic $\alpha \propto -\int \lb\bfv(t)\cdot\curl\bfv(t')\rb dt'
\sim -\tau_c \lb\bfv\cdot\curl\bfv\rb dt$
and $\beta\propto\int \lb\bfv(t)\cdot\bfv(t')\rb dt'
 \sim \tau_c \lb\bfv\cdot\bfv\rb dt$.  I will come back
to correcting the form for $\alpha$ in section 2.3 and 2.4.
because we really want a fully dynamic theory, that 
accounts for the dynamo coefficients'
dependence on  $\bfb$ and $\OB$.
 Only then can one fully address the fundamental
problem of mean field dynamo theory:  how does the growing
magnetic field affect the rate and saturation level of growth?
%The back-reaction on the dynamo coefficients to first
%order in $\overline {\bf B}$ and $\overline {\bf V}$ was calculated
%in Ref. 9, and Ref. 10 calculates $\alpha$ 
%to all orders in $\ob$ when mean field gradients are small.
%See also Refs. 11, 12, which find a catastrophically quenched
%alpha, but subject to the issues raised in Ref. 13.
%We will come back to the issue of catastrophic suppression is expected.

Substituting
(\ref{2.4aa}) into (\ref{2.4a}), 
gives the MFD equation
\be
{\partial\OB\over \partial t} = \nt \lrp{\OV\times \OB} 
+\curl(\alpha \bbB)-\curl(\beta +\la)\curl \OB. \lab{6}
\ee
The first term on the right is the non-controversial 
$``\Omega-{\rm effect}.''$
If one assumes $\overline {\bf V}$, 
$\alpha$ and $\beta$ to be independent of $\ob$, then 
(\ref{6}) 
can be solved as a linear eigenvalue
problem for the growing modes of $\OB$ 
in the Sun and other bodies.
%Boundary conditions play an important role in allowing net flux growth.
However a rapid growth of the fluctuating field necessarily accompanies the 
MFD in a turbulent medium.  
Its impact upon the growth of the mean
field, and the impact of the mean field itself on its own growth 
have been
controversial.  

The controversy arises because  
Lorentz forces from the growing magnetic field react back on 
the turbulent  motions driving the field growth  
\cite{kulsrud,cowling,piddington,vc,kitchatinov,ch,vainshtein}. 
%It is tricky to
%disentangle the backreaction from $\OB$ with that of the 
%fluctuating field. 
%{\bf***
It is useful to distinguish between fast MFD
action (also called  ``rapid'' MFD action)
and, slow MFD action (also called ``resistively limited'' MFD action).
Fast MFD action proceeds at 
growth rates which do not go to zero as the magnetic
Reynolds number $R_M\rightarrow \infty$, 
and maintain this property  even when the non-linear 
backreaction from the magnetic field is included.
Slow MFD action 
proceeds at rates that vanish as $R_M\rightarrow \infty$. 
I sometimes use ``resistively limited'' rather than ``slow'' 
because the former more explicitly describes the reduced action.
%}***

For galaxies and stars, conventional wisdom (which could be challenged
\cite{brannach} and see section 2.3)
presumes that rapid MFD action is necessary if the observed large
scale fields are to be wholly produced and sustained by the MFD, given
observed cycle periods and available time scales.
That this may not be the case is a separate issue from understanding
what the theory can actually provide. The latter is  the focus
herein.  If dynamos in stars and galaxies do operate fast, then 
we would like to understand how, in light of recent  
numerical and theoretical 
evidence for slow $\alpha^2$ dynamos in periodic boxes \cite{b2001,fb}. 
What are the differences between these dynamos
and real systems? where does the theory fail and where does
the theory succeed?

There are several issues which must be disentangled. 
First is the role of magnetic helicity conservation
in constraining dynamo theory.
In the steady state, such constraints
are strongly influenced by boundary conditions, so
one must be careful to understand the differences 
when applying idealized equations for closed systems 
to real systems with boundaries. 
Time-dependent dynamical constraints require helicity conservation 
to be supplemented in some way by the Navier-Stokes equation.
Then the fully dynamical evolution of 
$\alpha$ and the non-linear backreaction can be studied.

Two  directions emerge. One is to produce a 
time dependent dynamical non-linear
theory that fully agrees with  numerical simulations in periodic boxes.
This has recently been done \cite{fb}.
The second is to recognize that in real systems, there
is physics which has not yet been comprehensively studied. 
This includes shear \cite{vishniac,branshear}, boundary terms which alleviate
magnetic helicity conservation constraints, gravity, and a vertical
variation of the dynamo coefficients. 
It has proven difficult in the simplest generalizations,
to make the dynamo asymptotically fast \cite{brannach,branshear},
but this may not be needed in some applications (e.g. Galaxy)
if a kinematic phase lasts long enough. This will be discussed in detail.

In section 2 I discuss the role of magnetic helicity conservation
in dynamo theory. 
I first show that the escape of magnetic helicity through the boundaries
might play an important role in maintaining fast MFD action
during the steady sustenance phase of dynamos, 
and thus in interpreting quenching studies for this regime. 
I discuss the direct observational implications of the boundary
terms and the fact that stars and disks 
harbor active coronae.  
I then show that the time dependent mean field
dynamo is really a process by which magnetic helicity
gets transferred from small to  large scales by a non-local inverse cascade
and that the time-dependent dynamical  quenching in  
recent periodic box simulations can be understood in this framework 
\cite{fb}.  At late times, these box dynamos are 
resistively limited, depending sensitively
on the magnetic Reynolds number, however
the kinematic phase lasts quite awhile and this fact has very important
implications. In the last part of section 2, I discuss an unsolved puzzle 
that arises in the derivation of the successful 
dynamical quenching model.
%We also distinguish between helicity escape through outer 
%boundaries and helicity flow across a rotator's mid-plane.
In section 3, I summarize some key conclusions and pose open questions.

%One difference between the way
%large scale dynamos operate in galaxies and stars
%is that for the former, 
%the $\alpha$ effect and the $\Omega$ effect operate in the same
%volume. For stars however, 
%the $\alpha$ effect operates in the convection zone, whilst
%the $\Omega$ shear effect likely operates below, in the overshoot layer 
%[20].  

%{\bf 2. Dynamo Theory}

%The link is through   magnetic helicity, defined (Els\"asser
%1956) by
%\be
%H^M = \int_U \A\cdot \B\, d{\bf x} \lab{1}\ee
%where $\A$ is the vector potential of $\B=\nt \A$.  Field (1986) explains
%that $H^M$ is important in MHD  because under ideal conditions (vanishing
%resistivity) $H^M$ is conserved.  This has the consequence that in the $\ao$
%dynamo, in which kinetic helicity of the turbulence $\lb{\bv\cnt\bv}\rb$
%creates a large scale field carrying magnetic helicity, there has to be a
%compensating creation of small-scale field carrying 
%helicity of the opposite sign.

%\begin{eqnarray}
%  \dot{x}(0)&=&JH' (t,x)\\
%  x(0) &=& x(T)
%\end{eqnarray}
%with $H(t,\cdot)$ a convex function of $x$, going to $+\infty$ when
%$\left\|x\right\| \to \infty$.

\section{Magnetic Helicity Conservation and Dynamo Quenching}
%constraining $\lb\bfv\ts\bfb\rb, dynamo quenching, and coronal activity}

%\subsection{Role of Magnetic Helicity For Small Scale Field Spectrum}

%\subsection{Role of Magnetic Helicity For Mean Field Dynamo}

Although the MFD theory 
predates detailed studies of magnetohydrodynamic (MHD) 
turbulence, the MFD may be viewed as a framework for 
studying the inverse cascade of magnetic helicity.  
Whether this inverse cascade is primarily local (proceeding by
interactions of eddies/waves of nearby wave numbers) or non-local
(proceeding with a direct conversion of power from 
large to small wave numbers) is important to understand.
The simple MFD seems to be consistent with the latter \cite{b2001}.

From the numerical 
solution of approximate equations describing the spectra of energy and
helicity in MHD turbulence, Ref. \cite{pfl} showed 
that the $\alpha$ effect conserves magnetic
helicity, $H_M=\int({\bf A}\cdot{\bf B})d^3x$, 
by pumping a positive (negative) amount to scales $>l$ (the outer
scale of the turbulence) while pumping a negative (positive) amount to
scales $\ll l$. Magnetic 
energy at the large scale was identified with the $\OB$ of Ref. \cite{steenbeck}.  Thus,
dynamo action leading to an ever larger $\ob$, hence the creation of ever
more large scale helicity, can proceed as long as 
helicity of the opposite sign can be removed or dissipated.
More on this in section 2.3.  
%Recent simulations \cite{b2001} confirm this inverse cascade
%and the role of $H_M$ conservation.  
%Moreover, our recent recent semi-analytic theory for
%this process matches simulation results well.
%T%his analysis will be discussed in section 2.3.
Here I first derive the general magnetic helicity conservation
equations used to constrain the turbulent 
EMF, and investigate the implications of two steady
state cases in detail before considering the time
dependent case and interpretation of dynamo simulations. 

%The rate $H_M$ removal determines the rate of MFD action.
%We now discuss the role of boundaries for $H_M$ transport.

%The fate of small scale helicity is debated.
%According to the nonlinear solutions of Pouquet \etal (1976), it cascades
%to  large wave numbers where it is destroyed by Ohmic
%dissipation.  According to several authors (Cattaneo \& Hughes 1996,
%Gruzinov \& Diamond 1994, Seehafer 1994) the necessity for this process
%limits the buildup of large scale helicity, and hence, large scale magnetic
%fields. This would effectively eliminate the
%$\ao$ dynamo as a practical process for creating large scale magnetic
%fields in systems having a large magnetic Reynolds number $R_M$.

%\subsection{Magnetic helicity and the turbulent electromotive force}

%Blackman \& Field (2000) showed that 
%apparently conflicting
%simulations may correspondingly result from  
%whether or not the boundary conditions are periodic (Cattaneo \& Hughes 1996)
%or diffusive (Brandenburg \& Donner 1996) and whether there is significant
%scale separation. 
%To test astrophysically relevant dynamo issues, 
%periodic boundary conditions can be used if the averaging scale
%is significantly smaller than the size of the box (e.g. Pouquet et al. 1976;
%Meneguzzi et al 1981; Balsara \& Pouquet 1999).  This is not the case
%in Cattaneo \& Hughes (1996) who find a reduced $\alpha$, but for 
%the above reasons this reduction may be result from the boundary conditions
%rather than from a dynamical suppression.

Using Ohm's law for the electric field, 
\be
\bfE= {-c^{-1}\bfV \times \bfB +\eta {\bf J}} 
%= -c^{-1}\lb{\bfv\times \bfb}\rb-{\bbV}\ts\bbB+ \eta \bbJ, 
\label{ohm}
\ee
and averaging its dot product with $\bfB$, gives 
%[23]
\be
\lb{\bfE \cdot \bfB}\rb 
%&-{1 \over 2}(\partial_\mu {\tilde h}^\mu +\partial_\mu {\overline h}^\mu)  
=\bbE\cdot \bbB+\lb{\bfe\cdot \bfb}\rb 
= -c^{-1}\lb{\bfv\times \bfb}\rb\cdot \bbB+\eta \bbJ\cdot \bbB+
\lb{\bfe\cdot \bfb}\rb 
%\frac12
%\part_\mu { {\overline h}^\mu},
\lab{mon21}
\ee
%where 
%\be
%\bbE= \lb{-c^{-1}\bfV \times \bfB +\eta {\bf J}}\rb 
%= -c^{-1}\lb{\bfv\times \bfb}\rb-{\bbV}\ts\bbB+ \eta \bbJ \lab{mon6}
%\ee
%for the case $\bbV=0$, 
where ${{\bf J}}$ is the current
density.
%where we have used (\ref{mon6}).
%\beq
%\bbE \cdot \bbB= -c^{-1}\lra{\bfv\times \bfb}\cdot \bbB+\eta \bbJ \cdot
%\bbB\;, \lab{mon7}
%\ee
%which, from (\ref{2EMF}), shows that $\bbE\cdot \bbB$ is related to
%$\alpha$.  
A second expression for $\lb{\bf E}\cdot {\bf B}\rb$ also follows from 
Ohm's law without first splitting into mean and fluctuating components,
that is 
\be
\lb{\bfE\cdot \bfB}\rb 
%= \lb{-c^{-1} (\bfV\times \bfB )\cdot \bfB + \eta
=\eta \lra{\bfJ\cdot \bfB} = \eta \bbJ\cdot \bbB+\eta
\lb{\bfj \cdot \bfb}\rb =\eta\bbJ\cdot \bbB 
+c^{-1} \nu_M \lra{\bfb\cdot\nabla\times \bfb}.
\lab{mon22}
\ee
Using  
(\ref{mon22}) and  (\ref{mon21}), we have 
\be
-c^{-1} \lb{\bfv\times \bfb}\rb \cdot \bbB 
 = c^{-1} \la \lra{\bfb\cdot
\nb\times \bfb } -\lb{\bf e}\cdot {\bf b}\rb, 
\lab{mon23}
\ee
which can be used to constrain  $\lb\bfv\ts\bfb\rb$ in the mean field
theory.

%\subsection{Necessity of magnetic helicity escape}

Now consider ${\bf E}$ in terms of the vector 
and scalar potentials ${\bf A}$ and $\Phi$:
\begin{equation}
\bfE=-\nabla \Phi -(1/c)\partial_t {\bf A}.
\label{1}
\end{equation}
Dotting with $\bfB=\curl{\bf A}$ we have 
\begin{equation}
\bfE\cdot \bfB=- \nabla \Phi \cdot \bfB 
-(1/c){\bf B}\cdot \partial_t{\bf A}.
\label{02}
\ee
After straightforward algebraic manipulation and application of Maxwell's
equations, 
%(R\"adler 1980) 
this equation implies
\beq
\bfE\cdot \bfB
=- (1/2)\nabla\cdot   \Phi\bfB 
+(1/2)\nabla \cdot({\bf A}\ts {\bf E}) \nn\\
-(1/2c)\partial_t( 
{\bf A}\cdot \bfB) 
%=-(1/2c)\partial_t \int_U
%{\bf A}\cdot \bfB\ d^3x 
%=-\int_U \partial_t {H}^{0}d^3x 
%+\int_U \partial_i {H}^{i}
%=\int_U\partial_{0}
%{H}^{0}d^3x
=(-1/2c)\partial_\mu{H}^{\mu}(\bfB)=\eta\bfJ\cdot\bfB,
\label{03a1}
\eeq
where 
\beq
H^{\mu}(\bfB)=(H_{0},H_i)
=[{\bf A}\cdot \bfB, c\Phi\bfB 
-c{\bf A}\ts {\bf E}]
\label{heldef}
\eeq
is the magnetic helicity density 4-vector \cite{fieldaip},
and the contraction has been done with the 4 x 4 matrix $\eta_{\mu\nu}$ where
$\eta_{\mu\nu}=0$ for $\mu \ne \nu$, $\eta_{\mu \nu}=1$ for $\mu=\nu=0$ and 
$\eta_{\mu\nu}=-1$ for $\mu=\nu >0$.
Taking the average of (\ref{03a1}) gives
\be
\partial_\mu{\overline H}^{\mu}({\bfB})=-2c\lb\bfE\cdot \bfB\rb=
-2c\bbE\cdot \bbB-2c\lb\bfe\cdot \bfb\rb=-2c\eta\lb\bfJ\cdot\bfB\rb
\label{03a2}.
\ee

If, instead of starting with the total $\bfE$ as in (\ref{1}), 
I start with $\bfe$ and then dot with $\bfb$ and average, 
the analagous derivation replaces (\ref{03a2}) by 
\be
\partial_\mu {\overline H}^{\mu}(\bfb)=
-2c \lb {\bfe}\cdot {\bfb}\rb,
%-2c\eta\lb\bfj\cdot\bfb\rb
\label{14}
\ee
where ${\overline H}^{\mu}(\bfb)$ indicates the average of 
${H}^{\mu}(\bfb)$. The latter is defined like (\ref{heldef}) but with
the corresponding fluctuating quantities replacing the total quantities.
Similarly, starting with $\bbE$ and dotting with $\bbB$, 
gives
\be
\partial_\mu H^\mu(\bbB) 
=-2c {\overline {\bf E}}\cdot {\bbB},
%=-2c\eta\lb\bbJ\cdot\bbB\rb,
\label{15}
\ee
where ${H}^{\mu}(\bbB)$ is defined as in (\ref{heldef}) but with
the corresponding mean quantities replacing the total quantities.

In the next three subsections, I consider the implications of
magnetic helicity conservation for 3 separate cases. 
In the first I assume a steady state and ignore boundary
terms.  In the second, 	I include
boundary terms but still demand a steady state.
The third is the case in which boundaries are ignored, but
a full time evolution of $\alpha$ is considered and 
a fully dynamical theory is obtained that agrees
with recent numerical simulations.
%\cite{fb} that agrees with
%recent numerical simulations of Ref. \cite{b2001}.

%{\bf****
\subsection{Case 1:
Homogeneous, Stationary Turbulence in Periodic Box}
%}***

Consider statistically stationary turbulence, where the 
averaging is over periodic boundaries. 
Then, the spatial divergence terms on the left of 
(\ref{03a2}) become surface integrals and vanish. 
%We have 
%\be
%\partial_t{\lb\bfA\cdot\bfB\rb}=-2c\lb\bfE\cdot \bfB\rb=
%-2c\bbE\cdot \bbB-2c\lb\bfe\cdot \bfb\rb=-2c\eta\lb\bfJ\cdot\bfB\rb
%\label{03a2a}.
%\ee
Discarding the spatial divergenece terms in (\ref{14}) gives
\be
\lb {\bfe}\cdot {\bfb}\rb=-{1\over 2c}\partial_t \lb{\bfa}\cdot {\bfb} \rb.
%+\eta\lb\bfj\cdot\bfb\rb.
\label{17}
\ee
In the steady state, (\ref{17}) vanishes
and eqn (\ref{mon23}) shows that $\lb\bfv\ts\bfb\rb$ is resistively limited. 
%the 
%$\lb\bfe\cdot\bfb\rb$ is resistively limited by the second term on the
%right.  

If a uniform 
$\OB$ is imposed over the periodic box, 
then $\OB$ cannot change with time, and has no gradients.
This is the case of Ref. \cite{ch}, which measures $\alpha$ but cannot
test for MFD action. 
%The right side of (\ref{17}) vanishes because
%not only are zeroth order turbulent correlations 
%(i.e. those computed to zeroth order in the mean field) 
%stationary, but all higher order corrections must also 
%be stationary.  
No mean quantity varies on long time scales. 
%(When expanded in powers of the mean field, the first  
%contribution to (\ref{17}) enters at second order [23].)
In this case, (\ref{mon23}) 
then implies 
\be
%\bbE\cdot\bbB=
 -c^{-1}\lb{\bfv\times \bfb}\rb \cdot \bbB 
=\alpha {\bbB}^2/c =  c^{-1}\la \lra{\bfb\cdot\nb\times \bfb },
\lab{mon23q}
\ee
where $\alpha=\alpha_{33}$ for a uniform field in the $z$ direction.
Rearranging  gives  
\beq
|\alpha| = \left|{ \la \lra{\bfb\cdot\nb\times \bfb }\over  \bbB^2}\right|.
\label{19}
\eeq
%= %
%\left|{ \lra{\bfb\cdot\nb\times \bfb } \tau_{ed}\over R_M \bbB^2/v^2}\right|,
%If we assume that both $b$ and $v$ follow an $k^{-5/3}$ Kolmogorov energy 
%spectrum,  which is more justifiable in the case of maximally forced
%helicity (see section 1),
We then have 
%(\ref{19}) to find 
\beq
|\alpha| \lsim \left|{{k_2 b_2^2} \tau_{2}\over R_{M,2}^{n} \bbB^2/v_2^2}\right|
\sim \left|{\alpha_0\over R_{M,2}^{n} \bbB^2/b_2^2}\right|
\label{answer}
\eeq
where $b_2$ and $v_2$ are the fluctuating field magnetic energy 
of the dominant energy containing eddies (which is the forcing
scale for both $v$ and $b$ for $f_h=1$ turbulence as described in
section 1), $k_2$
is the wavenumber for that scale (= the forcing scale),  and
$n=3/4$ if the current helicity is dominated by large wavenumber
and $n=1$ if it is dominated by small wavenumbers.
The quantity 
$\tau_2$ is the associated eddy turnover time
and $R_{M,2}=v_2/k_2 \nu_M$ is the
magnetic Reynolds number associated with the forcing scale.
Assuming a steady state, I used  $v_2\sim b_2$
 which is roughly consistent with numerical results.
Then assuming forcing with maximal kinetic
helicty, I replaced the numerator with $\alpha_0$, the maximum
possible value of a helical quantity of that dimension.

\subsubsection{Pouquet Correction and Connection
to Previous Studies:}

If we now take $\alpha$ to be of the form 
first derived by \cite{pfl}, (discussed further in section 2.3) 
in the context of a maximally helical, force free $\alpha^2$ dynamo
in a periodic box,
we have
\beq
\alpha=-(1/3)\tau (\lb\bfv\cdot\curl \bfv \rb - \lb\bfb\cdot\curl\bfb\rb),
\label{PFLalpha}
\eeq
where $\tau$ is the correlation time of the turbulence
at $k_2$. 
Consider the implications of this formula in the steady state.
If we use (\ref{17}) in (\ref{mon23}), 
and allow for a 
non-uniform mean field 
it is straightforward to show that 
$\alpha$ can be written 
\be \alpha= 
{\alpha_0+R_{M,2}\beta \lb{\bbB\cdot \curl \bbB}\rb/{B^2_{eq}}\over
1+ R_{M,2} \bB^2/B_{eq}^2}.
\label{21}
\ee
Note that brackets around the current helicity 
are present because we allow for the fact that the mean
field can have a scale smaller than the overall system (or box)
scale.  For example, the growth of the mean field at wavenumber
$k=1$ in a periodic box can occur when the $k=0$ field is zero.
Indeed, the limit of this equation for a mean field of zero curl
produces  exactly the result obtained numerically by
\cite{ch} result for uniform $\bbB$.
However, as is clear from this formula, it only emerges
when $\bbB$ is uniform and the system is in a steady state.

For steady-state but non-uniform $\bbB$ in the simplest
case of a maximally helical $\alpha^2$ dynamo in periodic box with no shear,  
the field energy and large scale helicity 
sustenance depend on $\bbE\cdot \bbB$. If we 
assume for example that  
$\beta=\beta_0=$constant (unquenched, one extreme limit) and ignore mean velocities, 
and use (\ref{21}) 
\be
\begin{array}{r}
-\lb\bbE\cdot\bbB\rb= \alpha\lb\bB^2\rb 
- \beta \lb\bbB\cdot \curl \bbB\rb
={\alpha_0 +R_{M,2}\beta_0 \lb{\bbB\cdot \curl \bbB}\rb/B_{eq}^2\over
1+ R_{M,2} \bB^2/B_{eq}^2} \lb\bB^2\rb
-\beta_0 \lb\bbB\cdot\curl \bbB\rb
\\
={\alpha_0\over 1+ R_{M,2} \lb\bB^2\rb/B_{eq}^2}\lb\bB^2\rb 
-{\beta_0\over 1+ R_{M,2} \lb\bB^2\rb/B_{eq}^2}\lb\bbB\cdot \curl \bbB\rb,
\end{array}
\ee
where $B_{eq}=v_2$ in velocity units. 
Notice that choosing a constant $\beta=\beta_0$ 
in the steady state, leads to an EMF
that is resistively quenched, and is the same
as that with  an artificially imposed symmetric, resisitive
quenching of $\alpha$ and $\beta$. These forms of
$\alpha$ and $\beta$ are however misleading in the sense that
although their combination is consistent with helicity conservation
in the EMF, the division of $\alpha$ and $\beta$ in this way is 
not the division which was conistent with our initial assumption
of $\beta=\beta_0$ and so their forms are mere artifacts
\cite{bb}.  
Actually, in the saturated state, the current helicities of the large 
and small scale must be equal and opposite \cite{b2001,fb,bb}
(this follows from (\ref{03a2}) with no divergence terms). 
We then have $\lb\bbJ\cdot\bbB\rb\propto \alpha-\alpha_0$. 
This implies, for our assumed $\beta=\beta_0$, and for large $R_{M,2}$ that
\be \alpha= {\alpha_0 \over  1 + \lb \bB ^2 \rb /B_{eq}^2}.
\label{noq}
\ee
This  is the actual steady state form
of $\alpha$ when $\beta$ is unquenched. Note that there
is no $R_{M,2}$ depednence in the separate forms of
$\alpha$ or $\beta$ even though $\bbE\cdot \bbB$ is resitively
limited: in the steady state 
$\lb\bbE\cdot \bbB\rb$ must satisfy $\lb\bbE\cdot \bbB\rb
= \nu_M\bbJ\cdot bbB\rb$.  Ultimately, one can solve
for $\bB$ to see this demonstrated.

If we were to instead assume $\alpha \propto \beta$ from the outset,
one finds  
\be
\alpha= {\alpha_0 \over  1 + R_{M,2}(\alpha/\alpha_0 + 
\lb \bB^2\rb /B_{eq}^2 -1)}.
\label{apropb}
\ee

Later we will see a deeper implication of the appearance of
the current helicity in (\ref{mon23q}).   
%(Note that the ``Pouquet correction'' 
%to $\alpha$ \cite{pfl}, discussed further in section 2.3
%which is $\propto \lb\bfb\cdot\curl \bfb\rb$, enters to the same
%order in the mean field as the standard $\lb\bfv\cdot\curl\bfv\rb$ term 
%[10]. Thus we can consider the saturated value of 
%either to be representative of the 
%maximum kinematic magnitude of $\alpha_0$.)

%{\bf***
%The value of $\alpha$ in \pref{answer}
% is ``resistively limited,'' but the $R_{M,2}$ factor
%on the bottom results in our derivation simply as an a 
%priori implication of the chosen initial boundary conditions \cite{bf2000a}.
%No dynamics are directly involved here. 
%Stationary simulations with periodic boundaries and stationary turbulence   
%[18,22] are the most restrictive case, and their implications
%for time dependent dynamos or open systems can be misleading.

%}***

%{\bf****
\subsection{ Case 2: 
Inhomogeneous Turbulence, Finite Boundary Terms and Implications for
Coronal Activity in Steady State}
%}

Now  consider 
a system (e.g. Galaxy or Sun)  to have volume  $<<$  universal volume.
Integrating (\ref{03a1}) over all of space, $U$, then gives 
%\begin{equation}
%\int_U \bfE\cdot \bfB\ d^3x=-\int_U \nabla \Phi \cdot \bfB\ d^3x 
%-(1/c)\int_U{\bf B}\cdot \partial_t{\bf A}\ d^3x.
%\label{2}
%\end{equation}
%After straightforward algebraic manipulation, application of Maxwell's 
%(R\"adler ) and $\nabla\cdot \bfB=0$, this equation implies
\beq
\int_U \bfE\cdot \bfB\ d^3x=- (1/2)\int_U\nabla\cdot   \Phi\bfB\ d^3x 
+(1/2)\int_U\nabla\cdot( {\bf A}\ts {\bf E})\ d^3x \nn\\
-(1/2c)\partial_t\int_U{\bf A}\cdot \bfB\ d^3x 
%=-(1/2c)\partial_t \int_U
%{\bf A}\cdot \bfB\ d^3x 
%=-\int_U \partial_t {H}^{0}d^3x 
%+\int_U \partial_i {H}^{i}
%=\int_U\partial_{0}
%{H}^{0}d^3x
=-(1/2c)\partial_t{\EuScript H}(\bfB)=\int_U \eta {\bf J}\cdot \bfB d^3x  ,
\label{3a1}
\eeq
where the divergence integrals vanish when converted to 
surface terms at infinity.
I have defined  the global magnetic helicity  
%4-vector (Field 1986) 
\begin{equation} 
{\EuScript H}(\bfB)
\equiv\int_U{\bf A}\cdot \bfB\ d^3x,
%[H_{0},H_i]=[(1/2c){\bf A}\cdot \bfB, (1/2)\nabla\cdot  ( \Phi\bfB) 
%-(1/2)\nabla \cdot ({\bf A}\ts {\bf E})].  
\label{3aa}
\end{equation}
where  $U$ allows for scales much larger than the mean field scales.
It is easy to show that a parallel argument 
%using (\ref{1}), (\ref{02}), and (\ref{3a1}) 
for the mean and fluctuating fields leads to 
\be
\partial_t{\EuScript H}(\bbB)=
\partial_t\int_U\bbA\cdot\bbB\ d^3x
=-2c\int_U\bbE\cdot\bbB\ d^3x
\label{3aaa}
\ee
and
\be
\partial_t{\overline{\EuScript H}}(\bfb)=
\partial_t\int_U
\lb{\bf a}\cdot{\bf b}\rb\ d^3x
=-2c\int_U\lb\bfe\cdot\bfb\rb\ d^3x=-2c\int_U\bfe\cdot\bfb\ d^3x
=\partial_t{{\EuScript H}}(\bfb),
\label{3aab}
\ee
where the penultimate equality in (\ref{3aab}) follows from 
redundancy of averages.
%the 
%volume integral amounts to averaging over a scale larger than
%the mean scale.  
%The brackets and the overbar is taken
%to mean averaging over scale $\le$ that of $U$.
%If we take the average of (\ref{3a1})

%we then have 
%\be
%\partial_t{\overline{\EuScript H}}({\bf B})=
%\partial_t{\EuScript H}(\bbB)+\partial_t{{\bar {\EuScript H}}}
%(\bfb)=-2c
%\int_U \bbE\cdot \bbB\ d^3x-2c\int_U \bfe\cdot \bfb\ d^3x\simeq 0.
%\label{3ab}
%\ee
%Eqns. (\ref{3aaa}), (\ref{3aab}), and (\ref{3ab}) 
%show that a finite value of (\ref{mon23}) implies finite equal and opposite
%rates of change of global magnetic helicity from
%the small and large scale fields.
%Note that the time rate of change of the global helicities 
%are gauge invariant quantities.
%The exact same derivation carries through for both the 
%fluctuating 
%[h_{0},h_i]=[(1/2c){\bf  a}\cdot \bfb, (1/2)\nabla\cdot  ( \phi\bfb) 
%-(1/2)\nabla \cdot ({\bf a}\ts {\bf e})].  
%\be
%\lb{\bf E}\cdot {\bf B}\rb = 
%{\overline{\bf E}}\cdot \bbB +\lb\bfe\cdot\bfb\rb
%= {1\over 2}
%\partial_\mu{H}^{\mu}={1\over 2}
%\partial_\mu{\widetilde h}^{\mu}+{1\over 2}\partial_\mu{\overline h}^{\mu}\simeq 0,
%\label{ref1}
%\ee
I now split  (\ref{3aaa}) and (\ref{3aab})  
into contributions from inside and outside the rotator.
One must exercise caution in doing so because ${\EuScript H}$
is gauge invariant and physically meaningful only 
if the volume $U$ over which ${\EuScript H}$ is integrated is bounded by 
a magnetic surface (i.e. normal component of $\bfB$ vanishes at the surface), 
whereas the surface separating the 
outside from the inside of the rotator is not magnetic in general.
Ref. \cite{berger} shows how to construct a revised gauge invariant quantity  
called the relative magnetic helicity, 
\beq
{\EuScript H}_{R,i}({\bf B}_i)
= {\EuScript H}({\bfB}_i,{\bf P}_o)
-{\EuScript H}
({\bf P}_i,{\bf P}_o)  
\label{relhel}
\eeq
where the two arguments represent inside and outside the body respectively,
and $\bf P$ indicates a potential field.   
The relative helicity of the inside region is thus 
the difference between the actual helicity and the helicity
associated with a potential field inside that boundary. The use of
${\bf P}_i$ is not arbitrary in (\ref{relhel}), 
and is in fact the field configuration of lowest energy.  
While (\ref{relhel})
is insensitive to the choice of external field \cite{berger}, 
it is most convenient to take it to be a potential field as is 
done in (\ref{relhel}) symbolized by ${\bf P}_o$. 
The relative helicity of the outer region, ${\EuScript H}_{R,o}$,
is of the form (\ref{relhel}) but with the $o$'s and $i$'s reversed.
The ${\EuScript H}_R$
is invariant even if the boundary is not a magnetic surface.

The total global helicity, in a  
magnetically bounded volume divided into the sum of internal and 
external regions, $U=U_{i}+U_{e}$, satisfies  \cite{berger}
\be
{\EuScript H}(\bfB)={\EuScript H}_{R,o}(\bfB)+{\EuScript H}_{R,i}(\bfB),
\label{r4aaa}
\ee
when the boundary surfaces are planar or spherical.
This latter statement results from the
vanishing of an additional term associated with potential fields
that would otherwise appear in (\ref{r4aaa}).
Similar equations apply for $\bbB$ and $\bfb$, so (\ref{3aaa})
and (\ref{3aab}) can be written
\be
\partial_t{{\EuScript H}}(\bbB)
= \partial_t{{\EuScript H}}_{R,o}(\bbB) +\partial_t{{\EuScript H}}_{R,i}(\bbB),
\label{r4aab1}
\ee
and 
\be
\partial_t{{\EuScript H}}(\bfb)
=\partial_t{{\EuScript H}}_{R,o}(\bfb)
 +\partial_t{{\EuScript H}}_{R,i}(\bfb)
\label{r4aab2}
\ee
respectively. According to equation (62) of Ref. 25, 
\be
\partial_t{{\EuScript H}}_{R,i}(\bfB)=
-2c\int_{U_{i}}{\bfE\cdot\bfB} d^3x+2c\int_{D U_i}({\bfA}_p\ts {\bfE})\cdot d{\bf S},
\label{r4aac}
\ee
where ${\bfA}_p$ is the vector potential corresponding to a potential field ${\bf P}$ in $U_e$, and $DU_i$ indicates surface integration. 
Similarly,  
\be
\partial_t {{\EuScript H}}_{R,i}(\bbB)=
-2c\int_{U_{i}}{\bbE\cdot\bbB} d^3x
+2c\int_{D U_i}({\bbA}_p\ts 
{\bbE})\cdot d{\bf S}
\label{r4aad}
\ee
and
\be
\partial_t{{\EuScript H}}_{R,i}(\bfb)=
-2c\int_{U_{i}}{\bfe\cdot\bfb} d^3x+2c\int_{D U_i}{\bf a}_p\ts 
{\bfe}\cdot d{\bf S}.
\label{r4aae}
\ee
Note again that the above internal relative helicity time derivatives are 
both gauge invariant and independent of the field assumed in the
external region.  If we were considering the relative helicity of the
external region, that would be independent of the actual field in the internal
region.

Now if I take the average over a region $\le U_i$, I then
replace 
%(\ref{r4aad}) and 
(\ref{r4aae}) by
%\be
%\bbE\cdot\bbB={-1\over 2c}
%\partial_t{{\EuScript H}}_{R,i}(\bbB)+\div({\bbA}_p\ts 
%{\bbE}).
%\label{r4aadb}
%\ee
%and
\be
\lb{\bfe\cdot\bfb}\rb
=-{1\over 2c}
\partial_t{{\EuScript H}}_{R,i}(\bfb)+\lb\div({\bf a}_p\ts 
{\bfe})\rb,
\label{r4aaeb}
\ee
where the brackets indicate integrating over $U_i$ or smaller.
%If we demand a steady state, such that the first term
%on the right of (\ref{r4aadb}) and (\ref{r4aaeb}) vanish,
%and we assume 
We now see that even if the first term on the right of (\ref{r4aaeb}) vanishes,
$\lb{\bfe\cdot\bfb}\rb$ contributes a surface term to 
(\ref{mon23}) that need not vanish.  The turbulent 
EMF is not explicitly resistively limited as in the previous section.
since the surface term can dominate.
Thus in a steady state for $R_M>>1$, an outflow of magnetic helicity 
may enable fast MFD action.
Since the right of (\ref{03a2}) 
is small for large $R_M$,  the magnetic  helicity flux has contributions
from the small and large scale field.

%The above result highlights the role of total magnetic helicity 
%$H^M = \int_V \A\cdot \B\, d^3x$ (Els\"asser 1956), 
%where $V$ is a volume of integration,  
%and $\A$ is the vector potential.

%In short, $steady$ fast dynamo action  
%is possible in case 2 but not in case 1.
%This is consistent with current simulations.
%Case 1 is represented by  
%Refs. 21 and 22, which find that when the boundary
%conditions are periodic, dynamo action proceeds at resistively limited rates.
%In Ref. 27, action seems to proceed more rapidly, and this corresponds to a 
%case 2 simulation, though the restricted conditions require
%further analysis. 

%{\bf  *** 

%\subsection{Helicity Flow and Energy Flow to Coronae}

Note that the sign of the pseudo-scalar 
$\alpha$ coefficient changes across the mid-plane
of an astrophysical rotator.  This means that the magnetic
helicity generation in one hemisphere is of opposite
sign to that in the opposing hemisphere.  
One can then imagine that 
helicity flow could take place across the mid-plane; the loss of helicity
from say the top hemisphere into the bottom, 
acts in congruence with the loss of the opposite sign of
helicity from the bottom hemisphere into the top \cite{ji} allowing rapid MFD 
action.  Even if there is not strong coupling at the mid-plane
between the two hemispheres, an enhanced diffusion at the mid-plane 
side boundary of each turbulent/convective region could in principle
dissipate the helicity require to allow a rapid MFD and a rapid net flux
generation.

But there are reasons why a fast MFD in stars and galaxies  
would more likely involve escape of magnetic helicity out of the   
external boundaries in addition to merely a redistribution
across the mid-plane.  
First, buoyancy is important for both disks
and stars and the Coriolis force naturally produces
helical field structures in rising loops. 
Second note that the mean Galactic field appears to be quadrapole.
This already requires some diffusion at the upper and lower boundaries
of the disk \cite{parker}.  
Third, the solar cycle involves a polarity reverse of the mean dipole field 
and this requires the escape of magnetic fields out through the
solar surface.  
Fourth, note that for the Sun, the most successful dynamo models seem to be
interface models \cite{markiel} where diffusion just below the 
base of the convection zone is reduced, with no
alternative transport mechanism downward.
Finally, note that 
some aspects of solar magnetic helicity loss have been studied 
\cite{bergerruz} and observed \cite{rust,rustkumar}.

%\ni { 3. Relative magnetic helicity flow and associated energy flux} 
%\bigskip

%}****
%above is end bracket for bf

Let us  explore some implications of depositing 
relative magnetic helicity to 
a astrophysical corona.
%without addressing how the energy converts to particles or flows.
I assume that the rotator is in a steady state
over the time scale of interest, so the left sides of 
(\ref{r4aad}) and (\ref{r4aae}) vanish. 
For the sun, where the mean field flips sign 
every $\sim 11$ years, the steady state is
relevant for time scales less than this period,
but greater than the eddy turnover time ($\sim 5 \ts 10^4$ sec).
Beyond the $\sim 11$ year times scales, the mean large and small
scale relative helicity contributions need not separately
be steady and the left hand sides need not vanish.

The helicity supply rate, represented by the volume integrals
(second terms of Eqs. (\ref{r4aad}) and (\ref{r4aae})),
are then equal to the integrated flux of relative magnetic 
helicity through the surface of the rotator.  Moreover, from 
(\ref{03a2}), we see that the integrated flux of the large 
scale relative helicity, $\equiv {\EuScript F}_{R,i}(\bbB)$,
and the integrated flux of small scale relative helicity,
$\equiv {\EuScript F}_{R,i}(\bfb)$,
are equal and opposite.  We thus have
\be
{\EuScript F}_{R,i}(\bbB)=
-2c\int_{U_{i}}{\bbE\cdot\bbB} d^3x
=2c\int_{U_{i}}{\bfe\cdot\bfb} d^3x=
-{\EuScript F}_{R,i}(\bfb).
\label{r4aaf}
\ee
To evaluate this,  I use  (\ref{2.4a}) and (\ref{2.4aa}) to
find
\be
\bbE=-c^{-1}(\alpha\bbB-\beta\curl\bbB),
\label{r4aaf2}
\ee
throughout $U_i$.  
Thus  
\be
{\EuScript F}_{R,i}(\bbB)= -{\EuScript F}_{R,i}(\bfb)=
2 \int_{U_i}(\alpha\bbB^2-\beta\bbB\cdot\curl\bbB)d^3{\bf x}.
\label{r4aag}
\ee
This shows the relation between the 
equal and opposite large and small scale relative helicity
deposition rates and the dynamo coefficients.

Now the realizability of a helical magnetic
field requires its turbulent energy spectrum, $E^M_k$, to satisfy \cite{fplm}
\be
E^M_k(\bfb)\ge {1\over 8\pi} k|{{\EuScript H}}_k
(\bfb)|\; , \lab{10}
\ee
where ${{\EuScript H}}_k$ is the magnetic helicity at wavenumber $k$.
%The equality applies to force-free fields with $\nt \B=\pm k\bf B$. 
%Equation (\ref{10}) is really just 
%the recognition that the helicity is a signed 
%quantity whereas energy is unsigned.
The same argument also applies to the mean field energy spectrum, so that 
\be
E^M_k(\bbB)\ge {1\over 8\pi} k|{{\EuScript H}}_k(\bbB)|. 
\lab{10a}
\ee
If I assume that the time and spatial dependences are separable
in both $E^M$ and ${{\EuScript H}}$, then 
an estimate of power delivered to the corona can be derived.
I presume that the change in energy associated with the
helicity flow represents an outward deposition rather than an inward
deposition.  This needs to be specifically calculated for a given
rotator and dynamo, but since the source of magnetic energy is the rotator,
and buoyancy and reconnection interplay 
to transport magnetic energy outward, the assumption is reasonably motivated.

For  the contribution from the small scale field, 
we have 
%is associated with
%${\dot {\EuScript H}}(\bfb)$ 
\beq
\begin{array}{r}
\dot E^M(\bfb) = 
\int {\dot E}^M_k(\bfb) dk \ge {1\over 8\pi}\int k|
{\EuScript F}_{k,R,i}(\bfb)| \, dk
\ge  {k_{\rm min}\over 8\pi} \int | {{\EuScript F}}_{k,R,i}(\bfb)| \, dk
\nn\\ 
\ge {k_{\rm min}\over 8\pi} |{\EuScript F}_{R,i}({\bfb})|
={k_{\rm min}\over 8\pi} | {\EuScript F}_{R,i}(\bbB)|,
\end{array}
\lab{11}
\eeq  
where the last equality follows from the first equation in 
(\ref{r4aag}).
The last quantity is exactly the lower limit on 
$\dot E^M(\bbB)$.  Thus the sum of the lower limits on the total power 
delivered from large and small scales is 
${k_{\rm min}\over 8\pi} | {\EuScript F}_{R,i}(\bbB)|
+{k_{\rm min}\over 8\pi} | {\EuScript F}_{R,i}(\bfb)|
=2{k_{\rm min}\over 8\pi} | {{\EuScript F}}(\bbB)|$.  
Now for a mode to fit in the rotator, $k>k_{\rm min}=2\pi/h$, 
where $h$ is a characteristic scale height of the turbulent layer.  
Using (\ref{r4aag}), the total estimated energy delivered to the corona
(=the sum of the equal small and large scale contributions) is then
\beq
\dot E^M  \ge 2{k_{\rm min}\over 8\pi} |{\EuScript F}_{R,i}(\bfb)|
=2{k_{\rm min}\over 8\pi} |{\EuScript F}_{R,i}(\bbB)|
={{\rm V} \over h} \left| {\alpha\ob^2-\beta\bbB\cdot\curl\bbB}\right|_{ave},
\label{result}
\eeq
where $\rm V$ is the volume of the turbulent rotator and $ave$ indicates
volume averaged. I will 
assume that the two terms on the right of (\ref{result}) 
do not cancel, and use the first term of (\ref{result}) as representative.

%\ni { a. Solar Corona}

Working in this allowed time range for the Sun $(5\ts 10^{4}{\rm sec}<t< 11
{\rm yr}$) I apply 
(\ref{result}) 
to each Solar hemisphere. Using the first term as an
order of magnitude estimate gives 
\beq
{\dot E}^M \gsim  \left({2\pi R_\odot^2 \over 3}\right)\alpha \bbB^2=
10^{28} \left({R \over 7\ts 10^{10}{\rm cm}}\right)^2 
\left({\alpha \over 40{\rm cm/s}}\right)
\left({{\overline B} \over  150 {\rm G}}\right)^2 {\rm {erg\over s}}.
\label{sun}
\eeq
I have taken $\alpha\sim 40\,$cm~s$^{-1}$ (a low value of $\alpha$)
and  have presumed a field of 150G 
at a depth of $10^4$km beneath the solar surface in the convection zone,
which is in energy  equipartition with turbulent kinetic motions \cite{parker}.

As this energy is available for 
reconnection,  Alfv\'en waves, winds, and 
particle energization, we must compare this limit with the total of
downward heat conduction loss, radiative loss, and solar wind energy flux
in coronal holes, which cover $\sim 1/2$ the area of the Sun.  
According to \cite{withbroe}, this amounts to an approximately
steady activity of $2.5\times
10^{28}$erg~s$^{-1}$, about 3 times the predicted value of (\ref{sun}).
There is other evidence  
for deposition of magnetic energy and magnetic + current helicity
in the Sun \cite{rust,rustkumar,bf2000b}.

Active galactic nuclei (AGN) and the Galactic interstellar medium (ISM) 
represent other likely sites of mean field dynamos \cite{bf2000b}.
For the Galaxy,  ${\dot E}^M \gsim ({\pi R^2})\alpha \ob^2
\sim10^{40} ({R / 12{\rm kpc}})^2$ $\ts ({\alpha / 10^5{\rm cm/s}})
({\ob/ 5\ts 10^{-6}{\rm G}})^2 {\rm erg/s}$
in each hemisphere.
This is consistent with coronal energy input rates required by \cite{savage}
and \cite{rht}.
For AGN accretion disks, the deposition rate seems to be consistent
with what is required from X-ray observations.
Independent of the above, 
the most successful paradigm for X-ray luminosity in AGN 
is coronal dissipation of magnetic energy \cite{agn}.

So  boundary terms can potentially 
alleviate any helicity constraint and allow fast steady dynamo action.
But how this specifically happens and the specific boundary  physics
of a real system will need more study to see what field
wave numbers, if any, are preferentially shed.

\subsection{Case 3: Time-Dependent Dynamo Action and Dynamical Quenching 
in a Periodic Box}

For the question of actual field amplification and
time scales ultimately associated with cycle periods for a real dynamo, 
a time dependent, non-linear dynamical theory is required.
%to helicity conservation. 
%%helicity can be changed 
%injected  through the surface terms [26]. 
Here the time derivative term of $(\ref{03a2})$ important.
(Note also that time 
dependent dynamo like effects are also important for  magnetic 
field adjustment in Reverse Field Pinches \cite{ji}.) 

An important step forward was the work of Ref. \cite{pfl},  
based on the Eddy Damped Quasi-Normal Markovian spectral 
closure scheme. There an approximate set
of equations describing the evolution of magnetic and kinetic energy
and helicity was derived. Refs. \cite{maronblackman,b2001} performed 
numerical simulations of the process, and Ref. \cite{fb}
further simplified the equations of \cite{pfl}
 by considering a two-scale approach for fully helically
forced turbulence. In \cite{fb} it was assumed 
that the large scale field grows primarily 
on scale $k_1$ and the small scale turbulent field is
peaked at  $k_2 >> k_1$. The analytic model therein
is largely consistent with the forced helical turbulence, 
periodic box simulations of Refs. \cite{maronblackman,b2001}.
It should be mentioned however,
that  while the large scale field
dynamics are consistent with a single scale $k_1$ at all times, 
the wavenumber of the small scale peak seems to migrate from the resistive
scale to the forcing scale, in which case the two-scale approach
applies but with a migrating $k_2$ \cite{b2001,bb} unlike
that considered in \cite{fb}. More work is needed to understand
the migration of the small scale peak.  
For present purposes we ignore this complication which
does not effect the accuracy of the resulting  fits of the large scale field
growth.

The basic concept of the successful dynamical quenching 
model \cite{fb} for the $\alpha^2$ dynamo 
is that the growth of
the large scale field is the result of a segregation of magnetic helicity.
Magnetic helicity of one sign grows on the large scale, while the opposite
sign grows on the small scale, up to resistively limited conservation.
The growth of the small scale magnetic helicity also
grows current helicity which suppresses $\alpha$.
The two essential equations needed in this analysis are (1)
the equation for large scale magnetic helicity evolution (\ref{15})
and  (2) the equation for total magnetic helicity evolution (\ref{03a2}).
For more general dynamos with shear, the  
equations of the dynamical theory are (1) the vector equation for
the large scale magnetic field, and (2) 
the equation for total magnetic
helicity conservation.  In general, the paradigm that
emerges is that the total magnetic
helicity conservation acts as a supplementary dynamical equation
that is coupled to the evolution of the large scale field equation.

Ignoring the boundary terms in equation (\ref{15}). 
and subscript 1 and 2 to indicate mean and fluctuating scales,
and $M$ to indicate ``magnetic'' 
we then define $H^M\equiv H^0(\OB)/2$.
Thus (\ref{15}) implies
\beq
\partial_t H^M_1  =  2\alpha E^M_1 - 2(\la+\beta) {\overline {\bf J}}\cdot \OB
, 
\eeq
where where $E^M_1 = \OB^2/2$.
Now we replace spatial derivatives of the mean scale  with $k_1$,
and then note that the current helicity term on the right is related
to the magnetic helicity 
${\overline {\bf J}}\cdot \OB=k_1^2H^M_1$.
We then have  
%\doteq
\beq
\partial_t H^M_1  =  2\alpha E^M_1 - 2(\la+\beta) k^2_1 H^M_1\;.
\label{n40}
\eeq
Ref. \cite{fb} describes the conditions for which 
this is consistent with the analogous equation that arises
from the spectral treatment of \cite{pfl}.

The second equation we need is a re-write of (\ref{03a2}).
Since the left-most side is the sum of contribution
from large and small scales, as is the right-most side, we have 
in the two-scale approximation
\beq
\partial_t H^M_1  + \partial_t H^M_2  
=  - 2\la k^2_1 H^M_1- 2\la k^2_2 H^M_2.
\label{40p}
\eeq

We also need a prescription for $\alpha$ and  $\beta$.
A configuration space approach for $\alpha$ and its pitfalls
are described in the next section, but for now,
let us extract the dynamical form derived from \cite{pfl}.
For $\alpha$ this is
\beq
\alpha=-(1/3)\tau(\lb\bfv\cdot\curl \bfv \rb - \lb\bfb\cdot\curl\bfb\rb),
\label{PFLalpha}
\eeq
where $\tau$ is a correlation time of the turbulence
at $k_2$. In  externally forced simulations, the 
first term on the right of \pref{PFLalpha}
(the kinetic helicity)
is typically maintained at a fixed value. 
At early times, the second term on the right is small.  Thus 
the second term on the right (current helicity)
can be thought of as a  backreaction on the first term.
It arises from inclusion of the Navier-Stokes equation.
The idea that the second term represents a backreaction was
investigated by \cite{gd1,by} for a stationary system 
and a correction to $\alpha$ was derived. However, 
here I will derive the fully dynamical correction following \cite{fb}.
%kleeorin
First re-write \pref{PFLalpha} as
\beq
\alpha=\alpha_0+ \tau k_2 H^M_2 /3.
\label{n23}
\eeq
If we assume that the kinetic helicity is forced maximally, 
and take $\tau=2/k_2v_2$, then $\alpha_0= 2v_2/3$. 
%We also take $\beta_0=v_2/k_2$.

Unfortunately, there is not yet convergence on 
a rigorous prescription for $\beta$ in 3-D, but
I  will consider the two cases discussed earlier, 
$\beta=\beta_0\alpha/\alpha_0$
and $\beta=\beta_0\equiv v_2/k_2$=constant as examples.
In \cite{bb} other prescriptions are considered, including those
for which  $\beta$ is quenched but not resistively.
In general, the need for the appropriate form of $\beta$ is an important
input, and determines ultimately the form of saturated $\alpha$ 
in the $\alpha^2$
dynamo because the two are constrained through $\bbE\cdot \bbB$
as discussed at the end of section 2.1.
However, the particular 
form of $\beta$ is also less important for illustrating the role of
helicity conservation in the $\alpha^2$ dynamo than in dynamos
with shear as for the former 
since a range of choices are all relatively successful.  
I will show that 
the difference for large $k_2/k_1$ between  choices
of $\beta$ is really quite minimal. 
More prescriptions for $\beta$ are considered  in \cite{bb}.

%Ref. \cite{vainshtein} found both $\alpha$ and $\beta$ 
%to be suppressed similarly, but this was for waves.
The same type of formalism which leads to the
prescription for $\alpha$, produces a prescription for $\beta$
involving an approximate $sum$ of kinetic + magnetic energies times 
a correlation time, at least to first order sort of motivating
the case $\beta=\beta_0$.  More work
is ongoing and indeed for very strong mean fields since $\beta$ must at
least respond to the mean Lorentz force. 
The case $\beta=\beta_0\alpha/\alpha_0$ 
is motivated by the fact that at late times 
one empirical combination of formulae which fit \cite{b2001}
has this relation. 
% (Also 
%Ref. \cite{vainshtein} found both $\alpha$ and $\beta$ 
%to be suppressed similarly, but this was for waves.)

It is important to emphasize though, that 
as shown in section 2.1, a 
misleading degeneracy emerges in the steady-state (or near
steady state which amounts to the late time evolution) 
with respect to the choice of $\beta$. 
Again, the reason is that for $\alpha^2$ dynamos, it is really
the $\lb\bbE\cdot \bbB\rb$ that matters for the growth
of the large scale magnetic helicity, and thus the large scale
field. The main effect of the prescription for $\beta$ is
the saturation value of the field.

Following \cite{fb}, we need to solve 
 \pref{n40} and \pref{40p}. To do so, I rewrite them in dimensionless form.
Define the dimensionless magnetic helicities
 $h_1\equiv  2 H^M_1k_2/v_2^2$ and $h_2 \equiv 2 H^M_2 k_2/v_2^2$ and 
write time in units of $1/k_2v_2$. I also define  
$R_M\equiv (v_2/k_1)/\nu_M$.
(Note that this definition of $R_M$ is based on the forcing-scale
RMS velocity but on the large scale, $k_1^{-1}$.  We will later employ
the previously defined magnetic 
Reynolds number $R_{M,2}\equiv R_M (k_1/k_2)$.)

Using the above scalings we can replace 
(\ref{n40}) and (\ref{40p})
with dimensionless equations given by
\beq
\partial_t h_1 ={4\over 3}\left({k_1\over k_2}\right)h_1(1+h_2)-2h_1
\left[{k_1\over k_2 R_M}+{k_1^2\over k_2^2}(1+q_2h_2)\right]
\label{40b}
\eeq
and
\beq
\nn
\partial_t h_2 =-{4\over 3}\left({k_1\over k_2}\right)h_1(1+h_2)+2h_1
{k_1^2\over k_2^2}(1+q_2 h_2)
-{2\over R_M}{h_2k_2\over k_1},
\label{41b}
\eeq
where $q_2=0$ in the above equations corresponds to $\beta(t)=\beta_0=$constant.
and $q_2=1$ corresponds to $\beta(t)=\alpha(t)\beta_0/\alpha_0$.  
Solutions of these coupled equations are shown
in Figs. 3-8, which are taken from \cite{fb}. 
The key parameters are $k_2/k_1$, $R_M$, and $q_2$. 
In the figures, I have compared
these results to the empirical fits of numerical simulations 
in \cite{b2001}. I used  $h_1(t=0)=10^{-3}$,
but the sensitivity to $h_1(0)$ is only  
logarithmic (see \pref{tkin} below).
In Fig. 1,  $k_2/k_1=5$ was used, following B01, 
and in Fig. 2 $k_2/k_1=20$ was used.

In the figures, the solid lines represent numerical 
solutions to \pref{40b} and \pref{41b}, whereas 
the dotted lines represent the formula given 
in \cite{b2001}, which is an 
empirical fit to simulation 
data assuming that $\alpha$ and $\beta$ are prescribed according
to \pref{48} and \pref{49} below.
More explicitly, Ref. \cite{b2001} found that the growth of $\OB$ was
well described by the formula 
\beq
{B_1^2/B_{1,0}^2\over (1-B_1^2/B^2_{1,sat})^{1+{\alpha_0k_1-k_1^2\beta_0\over 
\nu_M k_1^2}}}=
e^{2(\alpha_0k_1-k_1^2\beta_0)t}, 
\label{axel1}
\eeq
where $B_{1,0}= B_1(t=0)$.
This can be rewritten using the notation above as 
a dimensionless equation for $t$ in units of $(k_2v_2)^{-1}$, namely 
\beq
t={k_2\over 2k_1}{ {\rm Ln}[(h_1/h_0)(1 - h_1 k_1^2/k_2^2)^{R_M(k_1/k_2-2/3)-1}]
\over {2 / 3} - {k_1/k_2}}.
\label{axel2}
\eeq
Note that \pref{axel1} and \pref{axel2} 
correspond to $\alpha$ and $\beta$ quenching of the form 
\beq
\alpha={\alpha_0\over  1+s_B B_1^2/v_2^2}
\label{48}
\eeq
and 
\beq
\beta={\beta_0\over  1+s_B B_1^2/v_2^2},
\label{49}
\eeq
where 
$s_B\sim R_M (k_1/k_2)(2/3-k_1/k_2)=R_{M,2} (2/3-k_1/k_2)$,
and $R_{M,2}\equiv v_2/k_2\nu_M$.
%After making a table of $t(h_1)$ from \pref{axel2}
%we can plot the points together with the solutions
%of  \pref{40b} and \pref{41b}
%% for the cases Rm=100, 1000 and k2/k1=5.
Eqns. \pref{48} and \pref{49} are derived from those in \cite{b2001} by
re-scaling Eq. (55) of \cite{b2001} with the notation herein.  
It can also be shown directly that, 
up to terms of order $1/R_M$,  
 \pref{axel2} is consistent 
with that derived by substituting \pref{48} and \pref{49} 
into \pref{40b} and solving for $t$.
Note that in contrast to the suggestion of \cite{b2001}, it is 
actually the forcing-scale magnetic Reynolds number, $R_{M,2}$, that plays 
a prominent role in these 
formulae.

The solutions of \pref{40b} and \pref{41b} 
are  interesting. Insight can be gained by their sum
\beq
\nn
\partial_t h_1 + \partial_t h_2 
= -{2\over R_M}\left({h_1k_1\over k_2} +{h_2k_2\over k_1}\right),
\label{41bb}
\eeq
which corresponds to 
\pref{40p}, the conservation of total magnetic helicity. 
If we make the 
astrophysically relevant assumption that $R_M >> 1$, the right hand side of
\pref{40p} is small for all $h_1$ and $h_2$. It follows that 
$\partial_t (h_1 + h_2)=0$ 
and for $h(t=0)=0$,  this implies $h_2 = - h_1$. 
In this period, we can self-consistently ignore $1/R_M$ in
\pref{40b}. If $q_2 = 1$, this phase ends 
when $h_2\rightarrow -1$, so that $h_1 \sim 1$. 
This is manifested in figure 5.
    
This kinematic phase precedes the asymptotic saturation of the dynamo 
investigated by other authors, in which all time derivatives
vanish exactly. For this to happen, the right hand side of \pref{41bb} must
vanish, which is equivalent to demanding that $h_2 = -(k_1/k_2)^2 h_1$. 
Since the right hand sides of \pref{40b} and \pref{41b} 
are proportional to $1 + h_2$ when terms
of order $1/R_M$ are neglected, their vanishing requires that $h_2 = -1$, 
and therefore, that $h_1 = (k_2/k_1)^2$. This is observed in figures 1 and 2.
The asymptotic saturation  (when the field growth ceases)
takes a time of order $t_{sat} \sim  R_M k_2/k_1$, 
which in astrophysics is often huge. Thus, although in principle it is correct
that $\alpha$ is resistively limited 
(as seen from the solutions in figures 6 \& 7)
as suggested by \cite{vc,ch,gd1,by}, this is 
less important than the fact that for a time $t_{kin} < R_M$ 
the kinematic value of $\alpha$ applies.
The time scale $t_{kin}$ here is given by
a few kinematic growth time scales for the $\alpha^2$ dynamo,
more specifically, 
\beq
t_{kin}\sim {\rm Ln}[1/h_1(0)] (k_2/k_1)/(4/3-2k_1/k_2).
\label{tkin}
\eeq
For $h_1(0)=0.001$, $k_2/k_1=5$, $t_{kin}\sim 37$, as seen in Fig 5.
 
Note that $t_{kin}$ is sensitive to $k_2/k_1$
and independent of $R_M$.  Figure 5 shows that there is significant
disagreement in this regime with \pref{48},
but this formula was used in \cite{b2001} only to model the  regime $t>R_M$, 
so the result is not unexpected. 
We can see from the solution  for $\alpha$ itself
that indeed the  solutions of \pref{40b} and \pref{41b} do match \pref{48} 
for  $t > R_M$ (figures 6 and 7).  
Figure 6 shows the difference in the $\alpha$ along with \pref{49}
for the two values $R_M=10^2$ and $R_M=10^3$. Notice again
the disagreement with the  formula \pref{48}
until $t=R_M$, and agreement afterward.  
This marks the time  at which the resistive term
on the right of \pref{40b} becomes competitive with the  
terms involving $(1+h_2)$. Asymptotic saturation 
does not occur until $t\sim t_{sat}=R_M k_2/k_1$ as described above.

%For $t>t_{kin}$, the current helicity
%term in $\alpha$, resulting from a growing $|h_2|$, begins to 
%significantly deplete the $\alpha$ effect.

Finally, note that $q_2=0$ corresponds to 
$\beta=\beta_0$.   
In general, this leads to a lower value of $h_1$ in the asymptotic 
saturation phase because 
this enforces zero saturation of $\beta$, whereas there is
still some saturation of $\alpha$ in this limit.
(Note that $q_2=0$ corresponds to the case of \cite{gd1} discussed 
in appendix B of \cite{fb}.)
For large $k_2/k_1$ the solutions of \pref{40b}
and \pref{41b} are insensitive to $q_2=0$ or $q_2=1$. This is because
the larger $k_2/k_1$, the smaller the influence of the
$q_2$ terms in \pref{40b} and \pref{41b}.  This is highlighted 
in figure 8 where the result for $q_2=0$ is plotted with the \cite{b2001}
fit. This suggests that for large-scale separation, the magnetic energy
saturation is  insensitive to the form of $\beta$ quenching. 
However, in real dynamos, magnetic flux and not just magnetic energy
may be needed, so the insensitivity can be misleading because
$\beta$ is needed to remove flux of the opposite sign.
From the low $k_2/k_1$ cases, it is clear that that $q_2=1$
is a better fit to the simulations of \cite{b2001}.

\begin{figure}[]
\begin{center}
\includegraphics[width=.8\textwidth]{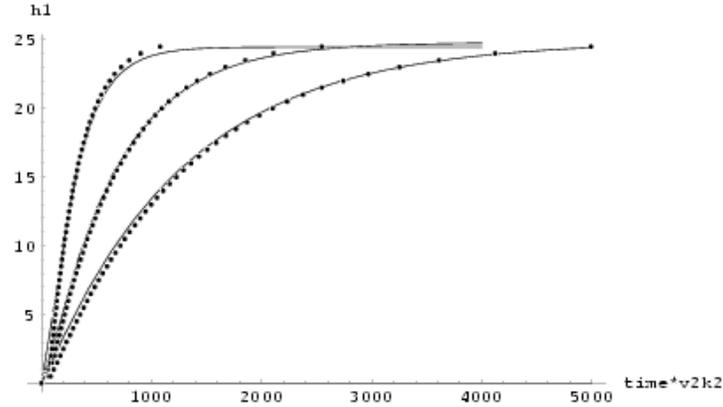}
\end{center}
\caption[]
{Solution for $h_1(t)$, $f_h=1$,$q_2=1$. Here $k_2/k_1 =5$ and
the three curves from left to right have $R_M=100,250,500$
respectively. The dots are plotted from the formula used to 
quasi-empirically fit the simulations in Ref. \cite{b2001}}
\label{H11002505005bran.ps}
\end{figure}

\begin{figure}[]
\begin{center}
\includegraphics[width=.8\textwidth]{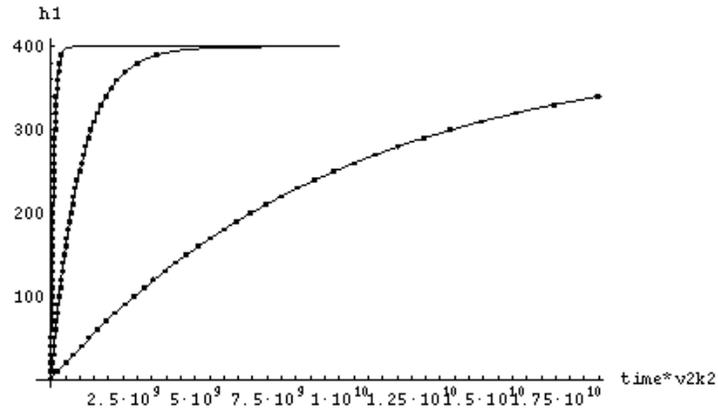}
\end{center}
\caption[]
{Solution for $h_1(t)$, $f_h=1$, $q_2=1$. Here $k_2/k_1 =20$ and
the three curves from left to right have $R_M=10^7,10^8,10^9$
respectively. The dots are plotted from the formula for used to quasi-empirically fit simulations of Ref. \cite{b2001}}
\label{H178920bran.ps}
\end{figure}

\begin{figure}[]
\begin{center}
\includegraphics[width=.8\textwidth]{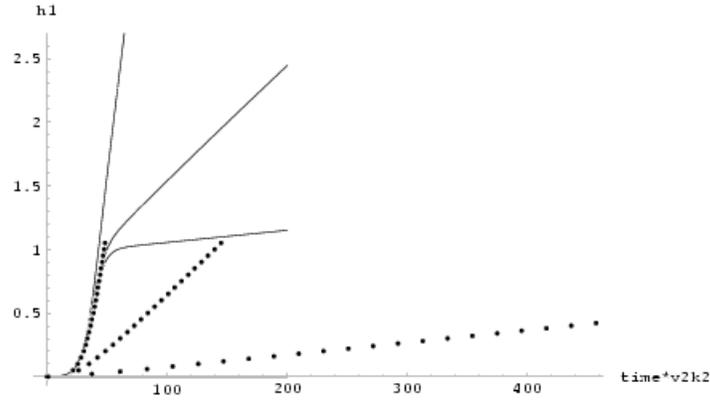}
\end{center}
\caption[]{The early-time solution for $h_1(t)$, $f_h=1$, $q_2=1$. Here  
for $k_2/k_1=5$, and $R_M=10^2,10^3, 10^4$ from left to right respectively.
Notice the significant departure from the 
formula of \cite{b2001} at these early times. For $t< t_{kin}$ there
is no dependence on $R_M$ and the growth proceeds kinematically.}
%The early time solution for $h_1(t)$, $f_h=1$, $q_2=1$. Here  
%for $k_2/k_1=5$, and $R_M=10^2,10^3, 10^9$ from left to right respectively.}
\label{h1R239k5e.ps}
\end{figure}

\begin{figure}[]
\begin{center}
\includegraphics[width=.8\textwidth]{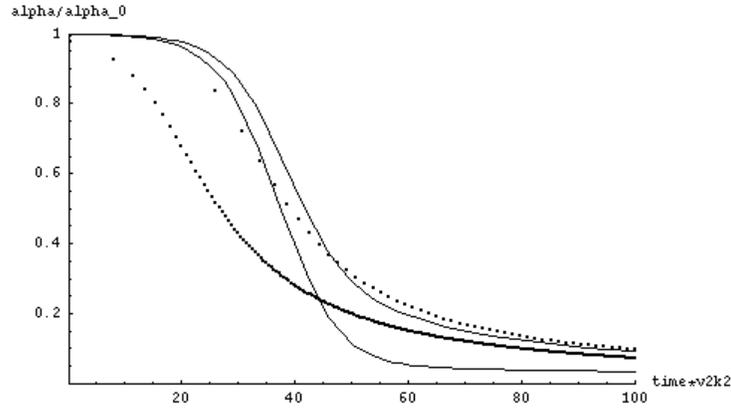}
\end{center}
\caption[]{Solution of $\alpha/\alpha_0(t)$ 
for $h_1(t)$, $f_h=1$, $q_2=0$. Here $k_2/k_1 =5$ and
the solid lines are the solutions to \pref{40b} and \pref{41b}
for $R_M=10^2$ (top curve)
and $R_M=10^3$ (bottom curve) respectively.
The top and bottom dotted curves are  from \pref{48}, interpreted
from Ref. \cite{b2001}. Notice the long kinematic
phase for the solutions, the overshoot, and the  convergence
with that of \pref{48} at $t=R_M$ for the $R_M=10^2$ case.}
\label{alpha235brane.ps}
\end{figure}

\begin{figure}[]
\begin{center}
\includegraphics[width=.8\textwidth]{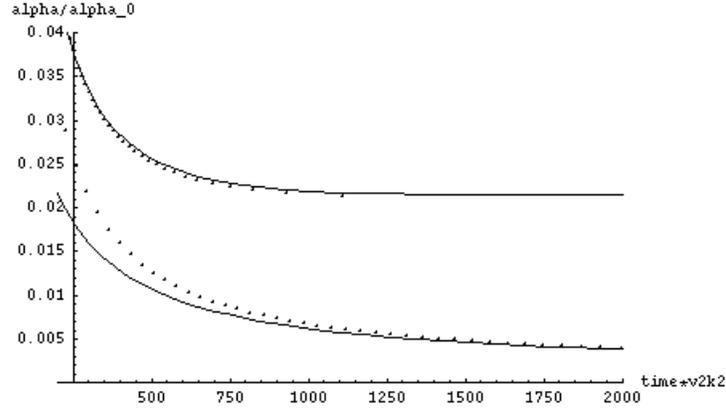}
\end{center}
\caption[]{This is the extension of  the previous figure for later times. Notice 
the convergence of the $R_M=10^3$ solution to \pref{48} near $t=R_M$.}
\label{alpha235bran.ps}
\end{figure}

\begin{figure}[]
\begin{center}
\includegraphics[width=.8\textwidth]{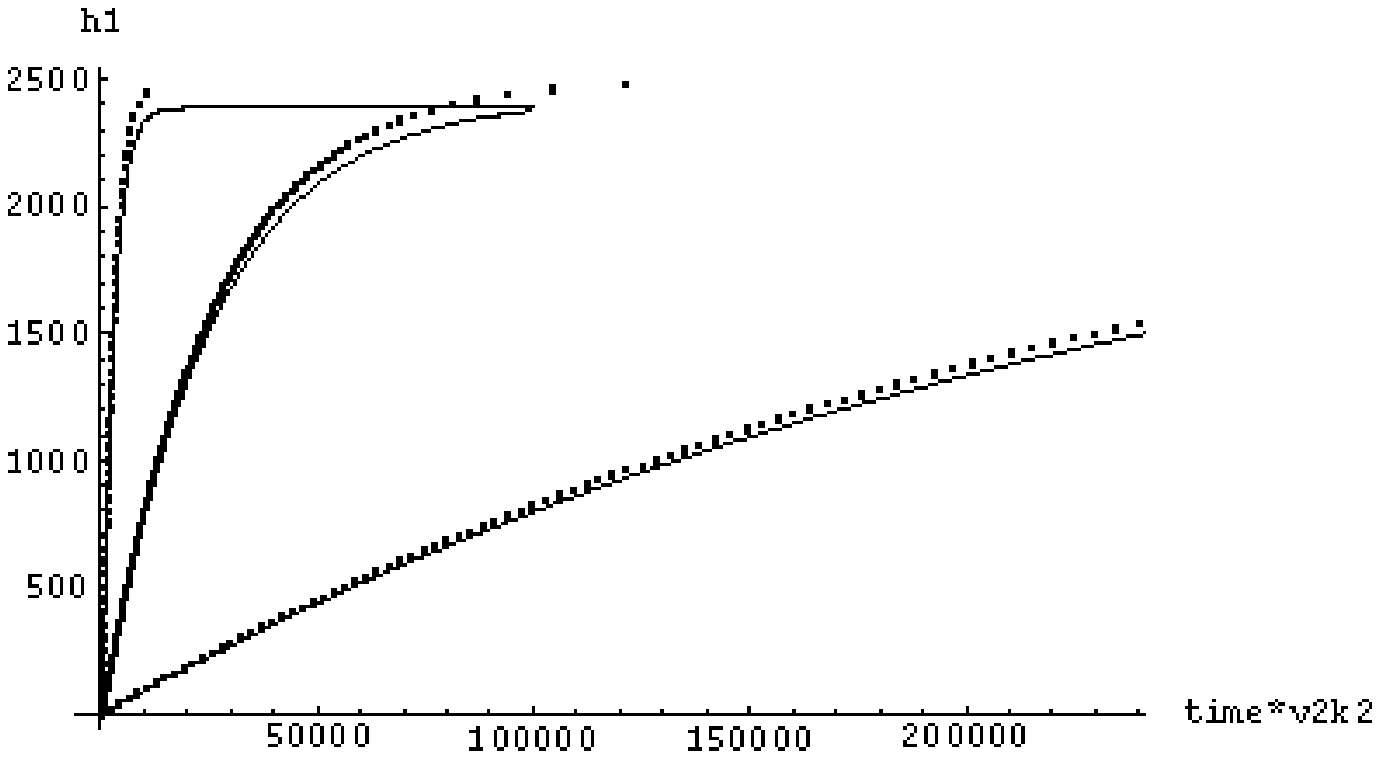} 
\end{center}
\caption[]{Solution for $h_1(t)$, $f_h=1$, $q_2=0$. Here $k_2/k_1 =50$ and
the three curves from left to right have $R_M=10^2,10^3,10^4$
respectively. The dotted lines are plotted from the formula 
used to quasi-empirically fit simulations of Ref. \cite{b2001}. For such 
large $k_2/k_1$ the fit to the data is only weakly sensitive to 
whether $q_2=1$ or $q_2=0$.}
\end{figure}

The physical picture of the quenching process just described
is this: helical turbulence is forced at $k_2$ ($=5$ in
\cite{b2001}), and kept approximately constant by forcing.  
Hence $\alpha_0 = -2\tau H^V_2/3 = {\it const}$.  
If $H_1^M$, the magnetic 
helicity at $k_1$ (which reaches $1$ here as a result of boundary
conditions), is initially small --- so that $|2k^2_1H^M_1/3|\ll |\alpha_0|$, \pref{n40} (or \pref{40b}) 
shows that it will be exponentially
amplified provided that the damping due to
$\beta +\la$ does not overcome the $\alpha$ effect.  
%It is important to
%note that turbulent diffusion can add to the
%effect of molecular diffusion by breaking up the field at $k_1$ and sending
%its helicity back to $k_2$.  
Initially, $\alpha=\alpha_0$, 
acting like a pump that  moves magnetic helicity from $k_2$ to $k_1$ 
and driving the dynamo.  This kinematic phase lasts
until $t_{kin}$ as given by \pref{tkin}. 
Eventually, the growing 
%From
%\pref{n44} and \pref{n53}, 
$H^M_1$
%, because of 
%current helicity conservation (see \pref{40p} or \pref{41b}), 
results in a  growing $H^M_2$ of opposite sign,
which reduces $\alpha$ through $H^C_2$.  
$R_M$-dependent quenching kicks in at $t=t_{kin}$, 
but it is not until $t=R_M$ that the asymptotic formulae 
\pref{48} and \pref{49} 
are appropriate.  Asymptotic saturation, defined by the time 
at which $B_1$ approaches 
its maximum possible value of $(k_2/k_1)^{1/2} v_2$,  
occurs at $t=t_{sat}=R_Mk_2/k_1$.
For $t\ge R_M$ the numerical solution 
of \pref{40b} and \pref{41b}, like the full numerical
simulations of \cite{b2001},
is  well fit by the $\alpha$ in \pref{48} with a corresponding $\beta$
of \pref{49}. The two-scale approach is also consistent with 
B01 in that magnetic helicity jumps from $k_2$ to $k_1$ without filling
in the intermediate wave numbers.

The emergence of the time scale $t_{kin}$ is 
interesting because it shows how 
one can misinterpret the implications of the
asymptotic quenching formula \pref{48} and \pref{49}.
These formulae are appropriate only for $t> R_M$. 
The large-scale field actually grows kinematically
up to a value $B_1=(k_1/k_2)^{1/2} v_2$  by $t=t_{kin}$
and ultimately up to $B_1\sim (k_2/k_1)^{1/2}v_2$ by $t=t_{sat}$.
For large $R_{M,2}$, 
these values of $B_1$ are  both 
much larger than the quantity $v_2/R_{M,2}^{1/2}$, which 
would have been inferred to be the saturation value if 
one assumed 
\pref{48} and \pref{49} were valid at all times.

Dynamical quenching or time-dependent approaches recognizing the current
helicity as a contributor to $\alpha$ 
have been discussed elsewhere \cite{zeldovich,kleeorinetal,kleeorinruz,kr}
(see also \cite{ji}), 
but here we have specifically linked the  PFL $\alpha$ correction to the 
helicity conservation in a simple two-scale approach. 
Other quenching studies for closed systems such as
%such as the work of Vainshtein \& Cattaneo (1992)
\cite{ch} and \cite{by} advocated
values of $\alpha$ which are resistively limited
and of a form in agreement with \pref{48} 
but with the assumption of a steady  $B_1$.
Assuming \pref{n23}, and using \pref{n40} and \pref{40p} in the steady
state, their formulae can be easily derived. 
However, one must also have a prescription for $\beta$.  
If $\beta$ is proportional to $\alpha$, then the resistively
limited formulae like \pref{apropb} emerges exactly,
which is indeed approximately consistent with 
\pref{48} and \pref{49} for large $R_M$.  If  $\beta(t)=\beta_0$, 
as in \cite{gd1}, then a formula for $\alpha$
$without$ resistively limited quenching \pref{noq} emerges.
On the other hand, fig. 8  shows that for large $k_2/k_1$, the 
dynamo quenching is largely insensitive to $\beta$.

Interestingly, if we cavalierly apply these results for the Galaxy
(by incorrectly ignoring the shear and assuming an $\alpha^2$ dynamo
that produces force free large scale fields),
and use $k_2/k_1=20$, $B_1(0)=10^{-9}$, 
$v_2=10$km/s,  and $k_2^{-1} \sim 100pc$, 
we would find the end of the kinematic
regime to be at $t=140$, or about $1.4\ts 10^9$yr. 
After this stage the field growth would proceed
very slow because of the large $R_M$, but the saturation field strength at this
time is  $B_1 \sim v_2/4.5$.  Thus quite
a large amplification can occur, even 
with an asymptotically slow dynamo.  However this is really
an academic exercise since for the Galaxy we need to consider
an $\alpha-\Omega$ dynamo, and the boundary terms. 

The generalized application of the principle that the magnetic helicity
conservations should supplement the mean field dynamo growth
equation to account for the backreaction is applied more
generally to dynamos with shear in \cite{bb}.

Finally, note that there is an important  puzzle, 
hidden in the derivations here 
and those of \cite{gd1,by} with regard to $\alpha$ quenching
that we discuss in the next section.

%Ref. \cite{b2001} showed numerically that in a periodic box, forcing with kinetic helicity.
%does indeed lead to growth of magnetic energy on the $k=1$ scale.
%The growth proceeds in two phases. 
%In the first phase, the field on the large scale and at the forcing
%scale both grow at the same rate.  In this phase,
%the, resistivity 

\subsection{Deriving $\alpha$ in Configuration Space: a Puzzle}

The two-scale dynamical theory of section 2.3 \cite{fb} for $\alpha$-quenching is 
appealing because it nicely couples 
the equations and concepts 
of magnetic helicity evolution 
to the current helicity contribution in $\alpha$, and fits simulation
data well.  The current helicity contribution was interpreted as a 
correction to the kinetic helicity contribution of kinematic theory.  
However, it depends
%the approach above,  and that of GD and BY, relies 
on  the current 
helicity contribution to $\alpha$ as presented in 
(\ref{PFLalpha}) being the
total current helicity associated with $k_2$.  To
see what we mean by {\it total} and to show the complication, we consider 
the derivation of the turbulent EMF in configuration space.

The turbulent EMF can be written in three different ways:
\beq
\begin{array}{rcl}
\lb\bfv\ts\bfb\rb = \lb\bfv(0)\ts\bfb(0)\rb + \int_0^t \lb\partial_{t'} \bfv(t')\ts\bfb(t')\rb dt' +
\int_0^t \lb\bfv(t')\ts\partial_{t'} \bfb(t')\rb dt'\\ 
\\
= \lb\bfv(t)\ts\bfb(0)\rb + \int_0^t \lb \bfv(t )\ts \partial_{t'}\bfb(t')\rb dt'\\ 
\\
= \lb\bfv(0)\ts\bfb(t)\rb + \int_0^t \lb \partial_{t'}\bfv(t')\ts\bfb(t)\rb dt' 
\end{array}
\label{c1}
\eeq
The three lines in \pref{c1} simply correspond to the 3 relevant 
ways of using the formula $f(t)= f(0) + \int^t_0 \partial_{t'}f(t')dt'$, where $g$ is an arbitrary
function of time. If I assume that 
$t>>0$, and that widely
separated turbulent quantities do not correlate, the first terms on the
right of the 2nd and 3rd lines respectively, can be dropped.
We then have
\beq
\begin{array}{rcl}
\lb\bfv\ts\bfb\rb 
=\int_0^t \lb \bfv(t )\ts \partial_{t'}\bfb(t')\rb dt' 
=\int_0^t \lb \partial_{t'}\bfv(t')\ts\bfb(t)\rb dt'. 
\end{array}
\label{c2}
\eeq
The  second term in \pref{c2}
as the standard textbook starting point \cite{moffatt,krause}
for evaluating the turbulent EMF for a kinematic dynamo, but here I
have not made any assumptions about the backreaction yet.

To illustrate the point, consider the  simple case in which $\nabla\OB=0$.
The equation for the small scale field is then
\begin{equation}
\begin{array}{r}\partial_{t'} \bfb = \OB\cdot\nabla\bfv + 
\bfb\cdot\nabla\bfv -\bfv\cdot\nabla\bfb -\curl\lb\bfv\ts\bfb\rb +
\nu_M\nabla^{2}\bfb.
\end{array}
\label{c3}
\end{equation}
The penultimate term goes away when included in \pref{c2}
and we ignore the last term.  The first terms on the right,
upon the assumption that the dominant contributions to correlations
are isotropic in $\bfv$,  gives the 
``textbook'' expression for $\alpha$ plus extra terms, that is  
\beq
\lb\bfv\ts\bfb\rb \simeq -{\OB\over 3}\int_0^t \lb\bfv(t)\cdot\curl\bfv(t')\rb dt'
+ Q (v^2 b).
\label{c4}
\eeq
The terms  symbolized by $Q(v^2 b)$ 
are typically ignored
using some version of the first order smoothing approximation.
This is of questionable validity, given that the small
scale field rapidly grows to exceed the mean field. 
We will come back to the relevance of these terms below.

Now if instead 	I use the last term of \pref{c2}
to expand the EMF, I must then invoke the Navier-Stokes equation for the
time derivative of the turbulent velocity 
\begin{equation}
\partial_{t} {\bfv}=-\bfv\cdot\nabla\bfv-\lb\bfv\cdot\nabla\bfv\rb
-\nabla p_{eff}
 +{\OB}\cdot\nabla{\bf b} + {\bf b}\cdot\nabla{\bf b}
-\lb {\bf b}\cdot\nabla{\bf b}\rb
+ \nu\nabla^{2}{\bf v}
+{\bf f}({\bf x}, t),
\label{c5}
\end{equation}
where $\bf f$ is a forcing function and $p_{eff}$ is the magnetic and
thermal pressure. Upon plugging this into \pref{c2},
the second and sixth terms  on the right vanish.
If we ignore the viscosity, and assume the dominant contribution
to correlations are isotropic, we then have 
\beq
\lb\bfv\ts\bfb\rb \simeq {\OB\over 3}\int_0^t \lb\bfb(t)\cdot\curl\bfb(t')\rb dt'
+ {\tilde Q} (v^2 b,fb)
\label{c6}
\eeq

Notice two things about \pref{c4} and \pref{c6}:
First they are equal to each other since they were derived from different
choices of the expansion of $\lb\bfv\ts\bfb\rb$.
Second, they do not cleanly include
the combination of the total residual helicity, required in
\pref{PFLalpha} when placed into \pref{n40} and \pref{40p} to derive \pref{40b} and
\pref{41b}.
The only way that  \pref{c4} and \pref{c6} can 
have the form of the desired relative helicity is if
$Q(v^2b)={\OB\over 3}\int_0^t \lb\bfb(t)\cdot\curl\bfb(t')\rb dt'$
and if 
${\tilde Q}(v^2b,fb)=-{\OB\over 3}\int_0^t \lb\bfv(t)\cdot\curl\bfv(t')\rb dt'$.
I have been unable  to prove that this is the case for the astrophysically
relevant weak $\bar B$ regime.

There is another approach to calculating the EMF
in configuration space that does  reveal a similar difference
of helicities as that in \pref{PFLalpha}, namely 
the approach of \cite{fbc}. 
%which led to the result in \pref{n8}.
%Indeed one is tempted by the similarity between \pref{n8} and \pref{n5},
%to make an identification between them, 
But $\bfv$ and $\bfb$ enter \pref{PFLalpha}
whereas  $\bfv^{(0)}$ and $\bfb^{(0)}$, the statistically isotropic
parts of $\bfv$ and $\bfb$,  enter \cite{fbc}. 
To see this more explicitly, I write 
\beq
\bfv=\bfv^{(0)}+\bfv^{(A)},
\label{c7}
\eeq
 where 
$A$ indicates an anisotropic contribution,
the result of the backreaction from $\OB$. Similarly, 
 \beq
\bfb=\bfb^{(0)}+\bfb^{(A)}.
\label{c8}
\eeq
(Even when $\bfb$ is the result
of stirring up an initial seed $\OB$, there is still a $\bfb^{(0)}$ which
is the statistically isotropic part of $\bfb$.)
We then assume that the statistics of the zeroth order
turbulent correlations are those of a homogeneous isotropic,
``known'' base state.
The goal is to express turbulent correlations in terms of the 
zeroth order quantities. It is sufficient to demonstrate the basic idea
invoked to all orders in $\OB$ in \cite{fbc} 
with that derived to linear order in \cite{bf99}
Noting that $\lb\bfv\ts\bfb\rb^{(0)}$ vanishes, the lowest
order contribution to the turbulent EMF is
\beq
\begin{array}{rcl}
\lb\bfv\ts\b\rb^{(1)} = 
\lb\bfv^{(0)}\ts\b^{(1)}\rb + \lb\bfv^{(1)}\ts\b^{(0)}\rb\\
= \int^t_0 \lb \bfv^{(0)}(t)\ts\partial_{t'} \bfb^{(1)}(t')\rb dt' + 
\int^t_0 \lb\partial_{t'} \bfv^{(1)}(t')\ts\bfb^{(0)}(t)\rb dt'
\end{array}
\label{c9}
\eeq
To linear order, using the induction equation for $\bfb^{(1)}$ and
the Navier-Stokes equation for $\bfv^{(1)}$, it can be
shown that by analogy to the derivations of \pref{c4} and \pref{c6},
(combined with a revised first order smoothing approximation
that assumes $|\bfb^{(1)}/\OB| < 1$)
\pref{c9} becomes
\beq
\lb\bfv\ts\bfb\rb^{(1)} = 
-{\OB\over 3}\left(\int^t_0 \lb\bfv^{(0)}(t)
\cdot\curl\bfv^{(0)}(t')\rb dt'-\int^t_0 \lb\bfb^{(0)}(t)
\cdot\curl\bfb^{(0)}(t')\rb dt'\right).
\label{c10}
\eeq
FBC showed that in the case of negligible mean field gradients,
it is still the zeroth order kinetic and current helicities which appear
most explicitly in $\alpha$, even to all orders in $\OB$.

It is clear that the zeroth order helicities are not necessarily 
equal to those constructed with the full turbulent quantities
since
%example, that 
% $\lb\bfv\cdot\curl\bfv\rb\ne \lb\bfv^{(0)}\cdot\curl\bfv^{(0)}\rb$
%results from the fact that  
\beq
\lb\bfv\cdot\curl\bfv\rb = 
\lb\bfv^{(0)}\cdot\curl\bfv^{(0)}\rb+
\lb\bfv^{(A)}\cdot\curl\bfv^{(0)}\rb+\lb\bfv^{(A)}\cdot\curl\bfv^{(0)}\rb
\label{c11}
\eeq
and similarly 
\beq
\lb\bfb\cdot\curl\bfb\rb = 
\lb\bfb^{(0)}\cdot\curl\bfb^{(0)}\rb+
\lb\bfb^{(A)}\cdot\curl\bfb^{(0)}\rb+\lb\bfb^{(A)}\cdot\curl\bfb^{(0)}\rb,
\label{c12}
\eeq
so the extra terms on the right must be dealt with.
One might ask however, if the first terms on the right of
\pref{c11} and \pref{c12} dominate, why can't we 
simply replace $\lb\bfb^{(0)}\cdot\curl\bfb^{(0)}\rb$ by
$\lb\bfb\cdot\curl\bfb\rb$ wherever the former occurs?
The reason is that the appropriate helicity which then enters $\alpha$  
is $\lb\bfb^{(0)}\cdot\curl\bfb^{(0)}\rb$ to lowest order. 
Then the $H_1^M$ entering on the left of \pref{n40} would
be second order. But then the magnetic helicity, $H_2^M$, entering 
 \pref{40p} would also be second order. 
Thus the current helicity entering $\alpha$ is zeroth order whereas
that entering the helicity conservation equation would
be second order. There is an ordering mismatch.
This is a problem because the success of the model of 2.3
depends on our being able to circumvent this ordering ambiguity
and presume that the current helicity entering \pref{PFLalpha} is exactly
$k_2^2$ times the $H_2^M$ entering \pref{40p}.

The procedure outlined to derive \pref{c10} 
and the subtlety just described with respect to ordering
is basically the  ``ordering ambiguity'' that was discussed in
\cite{bf99}.  There it was shown that 
Refs. \cite{gd1,by} effectively derived 
the form \pref{c9} rather than \pref{PFLalpha}
 by linearizing in terms of $\OB$ 
but did not identify that they had derived the zeroth order
contribution to $\alpha$.  Thus Refs. \cite{gd1,by} were actually
using a similar expansion to that of \cite{fbc}.  
The subsequent manipulations of \cite{gd1} and \cite{by} 
required  that they had derived $\alpha$ as a function of the full $\bfv$ and
$\bfb$, much like our manipulations in section 2.3.
The issue also arises subtly in the 
$k$ space derivation of PFL and is presently unresolved.

%A final point: the 
%assumption of a stationary and stable isotropic base state represented by
%zeroth order quantities such as $H^{M(0)}_2$ as invoked by
%\cite{fbc} may be realistic for a fixed $B_1$, but not  during the growth phase of a dynamo.  
%%Technically speaking therefore, the expressions for 
%%$\alpha$ given by  FBC and GD and BY, 
%%$\alpha$ that depends on the properties of the isotropic part of the
%%turbulence, denoted by superscript $(0)$.
%For example, in the $\alpha^2$ dynamo, such as that simulated in \cite{b2001}
%, instability drives the turbulence into a dynamical state
%with an ever-changing $B_1$.  This process
%drives $H^M_1(t)$, hence $H^M_2(t)$, away from that in the isotropic state
%that was the basis of the perturbative analysis of FBC and 
%implicitly of GD and BY.  This further highlights
%the importance of distinguishing zeroth order correlations
%from the full correlations.

\section{Conclusions and Open Questions}

\subsection{Small Scale Dynamo }

For non-helical turbulence, and for $Pr\ge 1$, current
simulations indicate that the magnetic field
piles up on the resistive scales $k>>k_f$ when forced externally 
\cite{maroncowley} in a periodic box.
The reason for this effect seems to be that the forcing scale
inputs shear directly into the small scale fields, so the power
on small scales is the result of cross field structure.
But for sufficiently helical turbulence, the spectrum changes:
the peak at the resistive scale migrates to the forcing scale (wavenumber
$k_f$) \cite{maronblackman}. 
Can we understand the migration of the small
scale field peak as a function of time? 
Will this picture survive future numerical testing?
How do the boundary conditions affect the results?

To explain the change in the small-scale spectrum 
from the non-helical to helical case, 
it is possible that what works for the large scale field may also
help understand what happens for the small scale field. 
The kinetic helicity input at $k_f$ also cascades
to higher wavenumbers, and so there is a source of helicity
at these wavenumbers.  Perhaps the change in shape of the small
scale spectrum might be modeled 
by a self-similar set of nested ``mean-field'' dynamos.
The principle is that for each small scale wavenumber 
$k_s > k_f $ there is a range of $k_f < k_l < k_s$ 
for which of inverse cascade field growth driven by the helical 
turbulence at $k_s$ can overcome the forward cascade of magnetic energy to 
$k>k_l$. An inverse cascade modeled in this way 
might account for the overall spectral shape change, but this is
presently just speculation.

\subsection{Large Scale Periodic Box $\alpha^2$ Dynamo}

The critical value of fractional helicity which determines
the migration of the small scale peak is exactly the same 
as that for which the kinematic $\alpha^2$ dynamo 
has a positive growth rate, and so  a large scale field
grows at $k_1 < k_f$ in concurrence with the migration of the
small scale peak to $k_f$ (see section 1 and 4).   
The rate of growth, the saturation level, and the dependence
on $R_M$ observed in periodic box simulations of fully non-linear
$\alpha^2$ dynamo seem to be well modeled by a two-scale dynamical 
quenching model of Ref. \cite{fb} and section 4. 
The dynamo $\alpha$ is the difference between the kinetic 
and current helicities associated with the small scale field, 
and the growth of the large scale field is associated with the growth
of the large scale magnetic helicity.
Because total magnetic helicity is conserved up to a resistive
term, the growth of large scale magnetic helicity also means 
growth of the opposite sign of small scale  magnetic helicity, and thus 
small scale current helicity.  This eventually suppresses
the growth of the large scale magnetic field by reducing $\alpha$.
The time dependent process is non-linear.
%It is also interesting that for Galaxy parameters,
%even the resistively limited growth of the large scale field
%is enough to produce a field strength which is approximately 1/20 of the
%equipartition value with the turbulence. 
There are some unresolved issues with this theory however:
1) the theory works well for maximally
helical forcing. How does it  generalize to arbitrary helical forcing?
2) How is the correction to $\alpha$ to be properly derived
in configuration space (see section 5.)?  The 
success of the theory is based on the formula
of $\alpha$ from \cite{pfl} which
leaves open the ordering ambiguity discussed in section 5. 
3) What is the appropriate theory for  $\beta$ quenching? 
The saturation value of the mean field in the $\alpha^2$ dynamo
depends somewhat on the choice of $\beta$ but the
ratio of large to small scale fields does not.
Dynamos with shear and their cycle periods
are sensitive to the form of $\beta$ \cite{bb}.
4) Along these lines, how does the dynamical quenching  theory
based on magnetic helicity conservation  
apply to the $\alpha-\Omega$ dynamo? This is work in progress \cite{bb}.
5) Do dynamos in astrophysics really need to be fast?
This is  another important reason to study $\beta$ quenching.
But the question should be revisited. Perhaps the kinematic
phase (see \pref{tkin}) can last long enough for substantial
amplification, even for asymptotically slow dynamos when no
cycle period is required. But one must also  consider the observed 
cycle periods like that of the sun which seem to be fast.
Perhaps highly anisotropic turbulence and small cross field structures
that can dissipate quickly might even allow fast 
cycle periods with through redefined $R_M$. See also \cite{brannach}. 

\subsection{Coronal Activity and Open Boundary Dynamos (OBD)}

In addition to these considerations, the role of boundary terms
needs to be considered. In a real system there
is gravity, rotation, shear, stratification and buoyancy.  It may not
be enough to just invoke arbitrarily open boundaries
to test the OBD but the physics at the boundary itself needs to
be studied (e.g. \cite{rafikov}).  Winds might be
essential.  Do mean field dynamos act in symbiosis with jets and winds?
We know that the sun is shedding
magnetic helicity. Does this enable the mean field dynamo to be fast for all
times? In astrophysical rotators, the $\alpha$ effect should depend
on height. How does this enter in the non-linear theory?
 
I showed above that the estimated energy deposition rates to maintain
fast mean field dynamo action are consistent with the coronal  + wind
power from the Sun, Galaxy and Seyfert I AGN  
all of these sources are natural sites for $\alpha-\Omega$ type dynamos.
The helical properties also seem to agree well in the solar case where they
can be observed \cite{rustkumar}.
The steady flow of magnetic energy into  coronae thus provides an interesting
connection between mean field dynamos and coronal dissipation 
in a range of sources.  A reasonably steady (over time scales
long compared to turbulent turnover time scales), active corona with
multi-scale helical structures, may provide a self-consistency check for 
fast dynamo production of magnetic field.

For the $\alpha^2$ case, figure 9 shows the effect of
including an additional loss term in \pref{41b},  
proportional to $h_2$,  on the growth of $h_1$.
The effect is to suppress the backreaction by taking away
excess small scale helicity and allowing the large scale 
field to grow  stronger.  
Understanding appropriate form of loss term for 
real  $\alpha-\Omega$ dynamos needs more study.

\subsection{New Diagram of MFD Operation is Needed}

Related to the boundary issue is 
the generation of magnetic flux inside of a rotator.
For the Galaxy, diffusion of $\OB$ at the boundary 
is required to maintain a quadrapole field structure with a net
flux inside the disk \cite{parker}. The total flux is conserved, but
to obtain a net flux inside the rotator where desired, 
the return flux must be shed from this region.
The surface diffusion of total magnetic field
for the Galaxy may be  difficult 
\cite{rafikov} and remains an open question.
(One alternative possibility is that the mean field diffuses
radially more efficiently at the top of the disk than at the mid-plane.) 
Note however, that it is $\OB$ which needs to diffuse, not 
necessarily the total field or the matter.  
For the Sun, the solar cycle also requires
diffusion of mean field through the boundary at the surface and/or mid-plane.

Since both flux generation inside the rotator, and helicity 
shedding can both appeal to boundary terms it is possible
that the two are related in the simplest $\alpha-\Omega$ dynamo.
Note that it was the growth of the
small scale  $current$ helicity $\lb\bfj\cdot\bfb\rb$ that
leads to the suppression of $\alpha$.
Thus one can ask directly if the textbook $\alpha-\Omega$ dynamo
has any way of getting rid of this quantity and at the same
time generating magnetic flux.

In fact there there is something missing when
one draws a standard picture of dynamo action
either for $\alpha^2$ or for $\alpha-\Omega$.  The latter
is shown in figure 10, but the implication holds also
shown for the $\alpha^2$ case.  The issue is that the small
scale loop that arises from the kinetic part of the $\alpha$ effect actually
induces the $same$ sign of the current helicity of the small
scale field as that of the large scale field that  it generates.
But I have discussed how the growth of the large scale field
should be accompanied by the $opposite$ sign of the small
scale helicities.  It is intriguing because in the kinematic
regime of the $\alpha^2$ dynamo, the small scale field does grow
first at the smallest scales, and it seems to take nearly until
$t_{kin}$ for the peak of the current helicity of opposite
sign to the large scale to move fully up to the forcing scale (Brandenburg and 
Maron, personal communication.) In this sense, figure 10 seems limited to
the kinematic regime:  it would be nice to see a more accurate 
picture that actually shows the generation of the small
scale current helicity of opposite sign to that of the large
scale and how it migrates to the forcing scale graphically.

\begin{figure}[]
\begin{center}
\includegraphics[width=.8\textwidth]{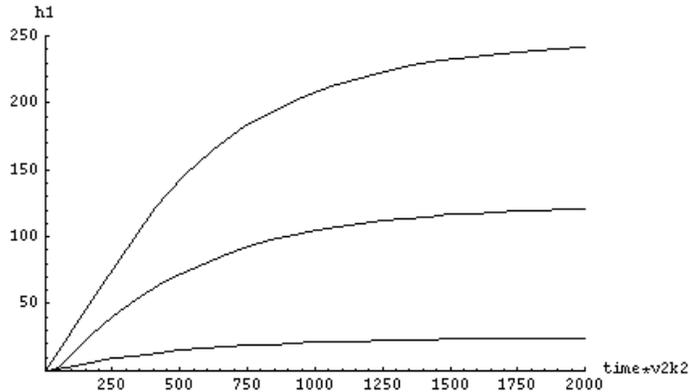}
\end{center}
\caption[]{The effect of a adding a loss term proportional
to $h_2$ in \pref{41b} on the growth of $h_1$ for three
different values of loss.  The loss term is
added by introducing multiplicative factors on the last 
term of \pref{41b}.   From top to bottom, these factors are
$10,5$ and $1$(=no loss), respectively. Here 
$R_M=200$ and $k_2/k_1=5$.}
\label{loss.ps}
%\label{dynamohelicity2.ps}
\end{figure}

\begin{figure}[]
\begin{center}
\includegraphics[width=.8\textwidth]{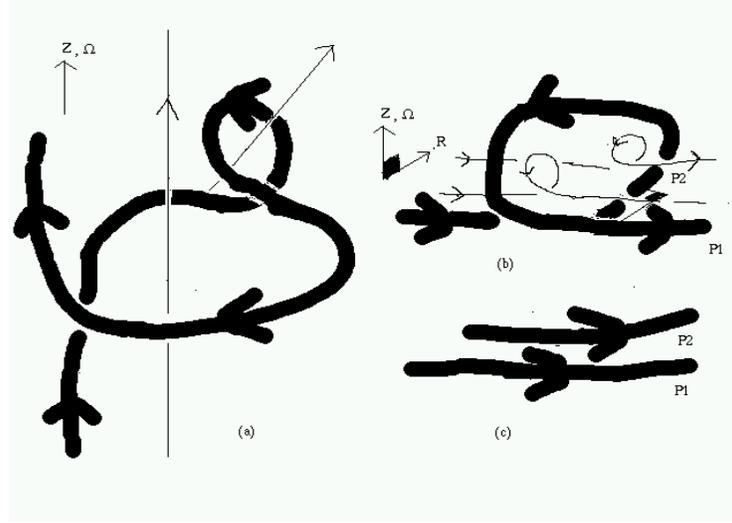}
\end{center}
\caption[]{The kinematic picture and its limitations. 
This is assumed to be a northern hemisphere so that kinetic
helicity is negative. (a) The large scale loop results
from shear and the small loop results from the $\alpha$ effect.
Note that the current helicity of the smaller loop has the 
same sign as that of the large scale loop. So the question 
arises: in a revised picture how can 
one incorporate the growth of magnetic and/or current helicity
of the opposite sign that accompanies growth of large scale
helicity? 
 (b) This just shows the
effect of multiple small scale loops as they sum together.
(c) the process of reinforcement of the toroidal field is shown.}
%\label{boundary.ps}
\label{dynamohelicity2.ps}
\end{figure}

Finally, note that 
it is in part the potential for flux generation that distinguishes
a mean field dynamo from the kind of dynamo which generates
large scale fields as a result of the magneto-rotational instability (MRI).
There, the field induces turbulence, which then generates a random
component of the field, which is subsequently sheared by the differential
rotation.  Magnetic energy grows exponentially, even on the largest
scale allowed, but there is no real flux generation here
and no need for input helicity.
Note however that for real accretion discs, helicity is undoubtedly
present since all the ingredients are there, density gradient, and
turbulence. Thus it is important to understand even in accretion
disks, what is the role of helical dynamos. This has not yet been
done exhaustively.  From this point of view, the mean field
dynamo formalism should still apply, with a  dynamical theory for
quenching, and the MRI may play the role as the source of turbulence.
Nevertheless more work is needed.

\subsection{What is the Role of Magnetic Reconnection?}

It is generally perceived that reconnection is important for large
scale dynamos but the precise way in which this is the case
is subtle. In general, the processes of 
reconnection can serve two roles: it changes the topology
of the field, and also dissipates magnetic energy.  
In the mean field dynamo formalism, the large scale field is degenerate
with respect to small scale topology. By definition, taking the
mean means smoothing out over the small scales. Thus a series
of disconnected loops could have the same mean field
as an undulating topologically connected field.  In the mean field
formalism therefore, the role of reconnection is not explicit.

Neither the mean field nor the fluctuating field are the actual field. 
If one really does want the mean field dynamo to result in a topologically
connected actual field, then reconnection would be important
for the topology. But it is important to asses the particular
application for when this is necessary: the large scale field
in the Galaxy is measured mainly by Faraday rotation which provides little
information about the actual field topology.  The role of reconnection
may only be one of ensuring that there is a turbulent
cascade which ensures that heat rather than the magnetic field
is sink of the kinetic energy.

Another unanswered  question relating to dynamos and reconnection
is: does reconnection play in the conservation of magnetic
helicity and in the migration of the peak of the small
scale magnetic energy from the resistive scale to the forcing
scale in helically forced turbulence?  

%\section{Equations}
%
%Please consult the file or printout of \emph{1readme}.$^*$ to find detailed
%instructions on the layout and style of mathematical elements.

%The following equation is the compiled result of using the command
%\verb|\vec| and using the newly defined command \verb|\umu|:
%\begin{equation}
%\vec{A}=\mu y = 50\; \umu {\mathrm m}\; .
%\end{equation}

\subsubsection*{Acknowledgments:}
Thanks to the organizers (E. Falgarone and T. Passot) 
of the Paris May 2001 workshop on MHD turbulence for an
excellent meeting. 
Thanks to G. Field, A. Brandenburg, and J. Maron for recent 
collaborations incorporated into the material herein. 
Thanks also to R. Kulsrud, B. Mattheaus, and A. 
Pouquet for stimulating discussions. Support from DOE grant 
DE-FG02-00ER54600 is acknowledged.

%INDEX%%%%%%%%%%%%%%%%%%%%%%%%%%%%%%%%%%%%%%%%%%%%%%%%%%%%%%%%%%%%%%%
% Please check with the editor of your book whether he plans to
% include a "mutual" subject index - if so, please code your entries
% in the standard syntax. For your own purposes you may print your
% "personal" index by using the following commands:
%
%\clearpage
%\addcontentsline{toc}{section}{Index}
%\flushbottom
%\printindex
%%%%%%%%%%%%%%%%%%%%%%%%%%%%%%%%%%%%%%%%%%%%%%%%%%%%%%%%%%%%%%%%%%%%%

\end{document}